\documentclass[prx,a4paper,twocolumn,superscriptaddress,groupedaddress,10pt]{revtex4-1}
\usepackage[T1]{fontenc}
\usepackage[ansinew]{inputenc}
\usepackage{amsmath,times}
\usepackage{mathrsfs}
\usepackage{amssymb}
\usepackage{graphicx}
\usepackage{xcolor,mathtools,bm}
\usepackage[colorlinks=true,linkcolor=purple,citecolor=green!10!blue]{hyperref}
\usepackage[capitalise]{cleveref}
\usepackage[top=.76in,bottom=.75in,right=.56in,left=.56in,heightrounded]{geometry}
\usepackage{xcolor}
\usepackage{amsthm}
\usepackage{mathtools}
\usepackage{physics}
\usepackage{graphicx,bm}
\usepackage[normalem]{ulem}
\usepackage{soul}
\usepackage{pgfplots}

\usepackage{cleveref}
\Crefformat{eqs}{Eq.~(#1#2#3)}

\definecolor{orange}{rgb}{1,0.5,0}

\definecolor{pink}{HTML}{f52c86}

\newcommand{\mycite}[1]{\cite{#1}}
\newcommand{\inlinecite}[1]{ref.\ \onlinecite{#1}}
\newcommand{\fig}[1]{Fig.~\ref{#1}}

\newcommand{\subfig}[2]{Fig.~\ref{#1}(#2)}
\newcommand{\subfigs}[3]{Fig.~\ref{#1}(#2) and (#3)}
\newcommand{\eq}[1]{eq.~(\ref{#1})}
\newcommand{\Eq}[1]{Eq.~(\ref{#1})}

\newcommand{\Sec}[1]{Sec.~\ref{#1}}
\newcommand{\app}[1]{app.~\ref{#1}}

\def\R{{\textbf{R}}}

\def\B{{\textbf{B}}}

\def\nb{{\boldsymbol{\nabla}}}

\def\para{{{\cal P}}}

\def\H{{{\cal H}}}
\def\P{{{\cal P}}}

\def\R{{\mathbf{R}}}
\def\r{{\mathbf{r}}}

\def\J{{\rm J}}
\def\a{{\rm a}}
\def\b{{\rm b}}
\def\c{{\rm c}}
\def\e{{\rm e}}
\def\vrho{{\varrho}}
\def\Z{{\cal Z}}
\def\S{{\textbf{S}}}

\newcommand{\rminus}{\mkern2mu\text{\scriptsize$-$}\mkern2mu}
\newcommand{\tminus}{\text{-}}

\def\R {{\mathbf{R}}}

\definecolor{crevise}{rgb}{.5,0.5,0.2}

\newcommand{\ignore}[1]{}

\def\k {{\mathbf{k}}}

\def\Sllp {{S_{ll'}}}
\def\Lllp {{\Lambda^{(ll')}}}
\def\Zllp {{\Z^{(l\leftrightarrow l')}}}

\ifdefined\added
\else
\newcommand{\added}[2][]{\textcolor{blue}{#2}\textsuperscript{\small\textcolor{red}{#1}}}
\definecolor{shadecolor}{RGB}{80,100,80}
\definecolor{pink}{RGB}{220,100,100}
\usepackage{marginnote}
\newcounter{mparcnt}

\newcommand{\nn}[2]{\langle #1,#2\rangle}
\newcommand{\nnn}[2]{\langle\langle #1,#2\rangle\rangle}
\fi

\begin{document}
\begin{titlepage}
\ifdefined\showplan
\fi
\end{titlepage}

\title{
Hidden Magnon Berry Curvature drives Vertical Magnon Transport
}
\author{Atul Rathor}
\email{atulrathor@bose.res.in}\affiliation{S. N. Bose National Centre for Basic Sciences, Block JD, Sector III, Salt Lake, Kolkata 700 106, India}
\author{Sahanawaj Akhtar}
\email{sahanawaj.akhtar@bose.res.in}
\affiliation{S. N. Bose National Centre for Basic Sciences, Block JD, Sector III, Salt Lake, Kolkata 700 106, India}

\author{Arijit Haldar}
\email{arijit.haldar@bose.res.in}
\affiliation{S. N. Bose National Centre for Basic Sciences, Block JD, Sector III, Salt Lake, Kolkata 700 106, India}
\date{\today}

\begin{abstract}
We predict an in-plane, or hidden, Berry curvature (BC) for magnons in electrically insulating quasi-2D magnets and demonstrate that the hidden magnon Berry curvature (HMBC) gives rise to a previously unrecognized form of vertical, out-of-plane, magnon transport.
Combining a semiclassical framework with Boltzmann transport theory, we show that the vertical magnon transport (VMT) currents respond both linearly and nonlinearly to the in-plane gradients of magnetic field and temperature. The linear transport coefficients are tied to the total hidden magnon BC, while the nonlinear (second-order) coefficients for the magnetic field and temperature gradients are determined by the hidden magnon BC dipole and the hidden extended magnon BC dipole, respectively. Using linear spin-wave theory, we find that the hidden magnon BC over the Brillouin zone is given by the expectation value of a pseudo-$\Z$ operator, representing vertical displacements, evaluated in the space of paraunitary matrices that diagonalize the magnon Hamiltonian. We estimate VMT in spin models of realistic magnets with ferro- and antiferromagnetic order, including the buckled honeycomb (BHC) lattice and bilayer Chromium trihalide (CrX$_3$; X = Cl, Br, I)  systems. 
In BHC, both linear and nonlinear VMT arise when time-reversal symmetry is broken by Dzyaloshinskii-Moriya interactions. In CrX$_3$ systems, the nonlinear coefficients dominate, while the linear responses vanish due to time-reversal symmetry. 
Both systems exhibit distinctive features across a broad range of temperatures and parameters. Therefore, our prediction of VMT and its characteristic signatures is directly testable in present-day magnonic experiments, especially in atomically thin, few-layered van der Waals magnets.

\end{abstract}
\maketitle

\section{Introduction}
Magnonic systems provide a promising alternative to electron-based devices \cite{Zutic2004Spintronics,Wolf2001Spintronics}. Magnons, which emerge as quasiparticles of magnetically ordered spins, are electrically neutral, making them efficient carriers of information and energy without incurring losses from Joule heating. In this regard, magnons offer a clear advantage over electrons, for which Joule heating losses are ubiquitous. However, this advantage also introduces difficulties: the charge neutrality of magnons prevents them from being excited by standard techniques used for electrons, particularly in the context of quantum transport. Understanding and leveraging quantum transport in both electronics and magnonics is essential, as it forms the foundation upon which present devices are built and future devices will be created. The key to unlocking the full potential of magnons, therefore, lies in harnessing and controlling their transport properties. The challenge is to impart magnons with transport characteristics comparable to those of electrons, despite the electrical neutrality of magnons. Significant progress has already been made towards realizing application-ready magnon-based alternatives to electron-based devices \cite{han2024magnonics,flebus2023recent}, and efforts continue to fully capture, and potentially surpass, the capabilities of electronic platforms within magnonics.

In recent years, two-dimensional (2D) and quasi-2D materials have emerged as promising platforms for exploring transport phenomena and exotic phases rooted in quantum geometry and band topology \cite{cayssol_topological_2021,Ma2010Abelian,
gianfrate2020measurement,wei_quantum_2023,schindler2018higher,Haldar.Higher.PRB,mook2021chiral}. For instance, transport arising from band topology, such as the topological Hall effect and related responses, has been observed in 2D electronic systems and remains under active investigation\cite{xiaoBerryPhaseEffects2010,du2021nonlinear}.  Even generalizations of transport phenomena mediated by real-space topological structures, such as skyrmions, have likewise been reported primarily in 2D materials\cite{yu2010real,Yang2020_Skyrmion_FGT_CoPd,Tong2018_MoireSkyrmions}. At a fundamental level, the band geometry and topology of these systems are encoded in the quantum geometric tensor, which quantifies the overlap of Bloch states at infinitesimally separated Bloch momenta \cite{zhu2025magnetic,resta2011insulating}. The real part of the tensor, termed as the quantum metric, measures the distance between Bloch-states, while the imaginary part is the familiar Berry-curvature (BC) that provides the key link between geometry and topology. The Brillouin zone (BZ) integral of the BC yields the Chern number, which is responsible for phenomena such as the topological Hall effect, while integrals involving momentum derivatives of the BC define the Berry curvature dipole (BCD), which drives nonlinear Hall responses \cite{du2021nonlinear,Sodemann2015quantum,Kondo2022nonlinear,Rathor2024spin}.

In particular, van der Waals (vdW) materials have proven especially successful for studying topology- and geometry-induced transport in electronic phases, due to the high degree of tunability offered by these structures \cite{Liu2016Van,Liu2022Synergistic,Qi2023Fabrication,Xu2024Van,Obaidulla2024Van}. vdW materials form a family of quasi-2D solids composed of a few atomically thin layers held together by weak van der Waals forces, enabling easy stacking and exfoliation. These features make vdW systems an ideal test bed for engineering diverse lattice and stacking geometries to tailor transport properties. The effects of quantum geometry on electronic transport in vdW materials are therefore under active exploration. Observations such as the spin Hall effect arising from tunable BCD \cite{Kormanyos2018Tunable}, giant nonlinear Hall responses in twisted bilayers \cite{He2020Giant}, and nonlinear anomalous Hall effects \cite{Kang2019Nonlinear} are a few examples from the rapidly growing list of transport phenomena in vdW systems.

While the majority of efforts have focused on electronic phases in vdW and related 2D materials, these platforms are increasingly being explored for magnonic applications \cite{Rahman2021recent,Shi2024recent,mi2023two,soriano2020magnetic,
Shishir2021dynamical,tian2019ferromagnetic,burch2018magnetism}. Experiments have reported the emergence of magnetism and various types of magnetic order in layered compounds such as FePS$_3$, FePSe$_3$, CrI$_3$,VI$_3$, Cr$_2$Ge$_2$Te$_6$, and related materials \cite{burch2018magnetism, Kim2020spin,Wildes2023spin,gong2017discovery,Basnet2024Understanding}. In particular, the chromium trihalide family of materials (CrX$_3$; X=Cl, Br, I) has become especially popular within the experimental community \cite{Huang2017layer}. Several studies have investigated how stacking arrangements and other tuning parameters influence the type of 2D magnetic order that develops in these systems\cite{chen2019direct,xu2022coexisting,zheng2018tunable,jang2019microscopic}. Most of these works have addressed the conditions necessary to achieve robust magnetic ordering in vdW materials. In comparison, the observation and characterization of magnon transport mechanisms in vdW systems remain at a nascent stage. While signatures of topological Hall--type magnon transport have been reported, experimental evidence of transport driven purely by band geometry is still lacking. On the theoretical side, however, proposals are rapidly emerging on how to harness geometry-induced transport phenomena in magnonic systems. For instance, recent studies have suggested the possibility of nonlinear thermal Hall effects arising from the magnon Berry-curvature dipole and its extended generalizations\cite{Rathor2024spin,mook2018taking}.

Since vdW materials are quasi-2D systems with a finite but minimal thickness along the out-of-plane ($z$) direction, it is natural to ask whether universal features of bulk 3D systems leave an imprint on such few-layered materials \cite{Hara2020current}. Indeed, it has been recently proposed that in electronic systems the full three-component Berry curvature, $\boldsymbol{\Omega} = (\Omega^{(x)}, \Omega^{(y)}, \Omega^{(z)})$, defined for bulk 3D crystals, may manifest in quasi-2D materials as in-plane components $\Omega^{(x)}$ and $\Omega^{(y)}$ in addition to the usual $\Omega^{(z)}$ component, which is always well defined in 2D systems \cite{bellissard1995noncommutative}. From this perspective, if one visualizes a hypothetical bulk 3D electronic system constructed by stacking weakly coupled copies of any given vdW structure along the $z$-direction, the in-plane BC components of the quasi-2D material can be understood as the limiting values of the bulk $\Omega^{(x, y)}$ components projected onto the $xy$-plane of the isolated vdW structure. As a direct consequence, these in-plane components vanish identically in perfectly flat, single-layer 2D electronic systems. For this reason, the additional in-plane BC contributions in few-layered electronic systems have been termed hidden Berry curvature, as they are absent in strictly 2D materials but can emerge in layered quasi-2D electronic structures.

This raises the central question: Does a dual of the hidden electronic BC exist for magnons, and more generally for bosons? If so, can this magnonic generalization of hidden BC lead to measurable properties that remain unexplored? In this paper, we show that a magnon dual, which we dub the hidden magnon BC (HMBC), can indeed be identified for magnon bands arising in quasi-2D magnets. The hidden magnon BC turns out to be the in-plane components of the generalized symplectic BC of bosonic excitations on lattices.  More importantly, we also find that a nonzero HMBC can drive vertical magnon transport (VMT) perpendicular to the plane of quasi-2D magnets, thereby providing a direct way to quantify the hidden geometry of magnons through transport measurements.

The key findings of our work are as follows -- \emph{First}, using a semiclassical magnon wave-packet description combined with the Boltzmann transport equation, we show that the vertical magnon current is explicitly linked to the $\Omega^{(x)}$ and $\Omega^{(y)}$ components of the BC and is generated in response to applied magnetic-field or temperature gradients. In the same analysis, we establish while the magnetic field can point along any direction, finite in-plane field gradients are necessary to produce VMT. \emph{Second}, expanding the vertical current to second order in field and temperature gradients, we discover: the linear magnetic-gradient coefficient is proportional to the BZ integral of the hidden BC weighted by the Bose-Einstein (BE) distribution, while the linear temperature-gradient coefficient is a similar integral weighted by an alternate function of the BE distribution. The second-order coefficients are integrals comprising another well-known bosonic distribution function and momentum derivatives of the hidden BC (for field gradients) or of the energy-weighted hidden BC (for temperature gradients). These integrals resemble the magnon BCD \cite{Rathor2024spin} and the extended magnon BCD \cite{Kondo2022nonlinear} originally defined for the BC component $\Omega^{(z)}$; thus, we name these previously unexplored coefficients as the hidden magnon BCD and the hidden extended magnon BCD, respectively. \emph{Third}, from a microscopic approach we show that the hidden components $\Omega^{(x,y)}$ can be obtained from a symplectic displacement operator -- named the pseudo-$\Z$ operator -- which represents displacement along the out-of-plane direction. Evaluating its expectation value using the single-magnon modes obtained from diagonalizing the magnon BdG Hamiltonian yields the hidden BC. \emph{Fourth}, we establish the symmetry requirements for obtaining VMT from the hidden magnon BC. Specifically, second-order VMT due to the hidden magnon BCD can arise only when at least one layer in the quasi-2D structure is non-identical, while linear-order VMT from HMBC \emph{additionally} requires time-reversal symmetry breaking in the single-magnon sector.

The remainder of this paper is organized as follows. In \Sec{Sec:VMT}, we derive the dependence of vertical magnon current on magnetic field and temperature gradients using a semiclassical framework and Boltzmann transport theory. In \Sec{Sec:Hidden_BC}, we develop a microscopic formulation of the hidden Berry curvature components in terms of the pseudo-$\Z$ operator. In \Sec{Sec:models}, we predict and characterize vertical magnon transport in realistic quasi-2D systems, specifically the buckled honeycomb lattice and bilayer chromium trihalides. We summarize  and discuss the implications of our work in \Sec{Sec:conclusion}.
\section{Vertical magnon transport}\label{Sec:VMT}
\begin{figure}
\includegraphics[width=.5\textwidth]{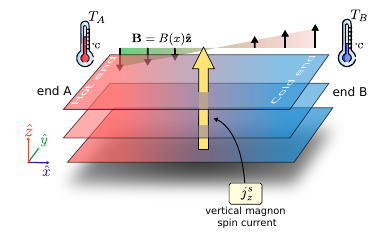}
\caption{\textbf{Vertical (out-of-plane) magnon current in quasi-2D systems}: A quasi-2D magnetic insulator consisting of a few atomically thin layers is placed on the $xy$-plane. The system is subjected, either separately or simultaneously, to a spatially varying magnetic field ${\B}=B(x)\hat{\textbf{z}}$ with an in-plane gradient and a thermal gradient along the $\hat{\textbf{x}}$ direction arising from a temperature differential $T_B-T_A$. The hidden magnon Berry curvatures of the quasi-2D magnet give rise to a vertical (out-of-plane) magnon current.}\label{fig:setup}
\end{figure}

We consider a magnetic insulator, subjected to a spatially varying magnetic field ($\B$) and a temperature ($T$) gradient, applied either independently or simultaneously, as shown in \fig{fig:setup}. We further assume the insulator is described by an effective spin Hamiltonian on a lattice having a ground state hosting a stable collinear magnetic order. Therefore, the collective low-energy excitations of the lattice spin Hamiltonian constitute magnons that carry spin angular momentum and hence can interact with the applied magnetic field through the Zeeman effect \cite{aharonov1984topological,mook2018taking}. The dynamics of magnons at the semi-classical level can be described by constructing a magnon wave packet obeying the following  equations \cite{xiaoBerryPhaseEffects2010}:
\begin{subequations}
\begin{align}
\dot{\r}=&~\hbar^{-1}\partial_\k(E_{n, \k}-\textbf{}\bm{\mu}_{n, \k}\cdot\B)-\dot{\k}\times \bm{\Omega}_{n, \k},\label{eqm1}\\
\hbar \,\dot{\k}=&-\nabla(\bm{\mu}_{n, \k}\cdot\B),\label{eqm2}
\end{align}
\end{subequations}
where ${\r}$ and $\hbar\,{\k}$ are the  position and momentum of the center of the  wave-packet, while $E_{n, \k}$  and $\bm{\Omega}_{n, \k}=(\Omega^{(x)}_{n, \k},\, \Omega^{(y)}_{n, \k}, \,\Omega^{(z)}_{n, \k})$ are the band dispersion and Berry-curvature of the $n$-th magnon Bloch band, respectively. The vector $\bm{\mu}_{n, \k}$ indicates the magnetic moment, and hence spin-angular momentum, carried by a magnon, and can be obtained by solving for the magnon band dispersion in the presence of a $\B$-field and calculating $\bm{\mu}_{n,\k}=-\partial E_{n,\k}(\B)/\partial \B|_{\B =\,\bm{0}}$ \cite{Neumann2020orbital}. \Eq{eqm1} shows that the wavepacket velocity is not only determined by gradient forces generated due to the energetics in system (see first term) but also by the anomalous velocity term $\dot{\k}\times\bm{\Omega}_{n, \k}$ \mycite{xiaoBerryPhaseEffects2010} arising from the BC of the magnon bands.

At a qualitative level, \eq{eqm1} is sufficient to hint towards the possibility of vertical magnon current from hidden BC in quasi-2D materials.
As discussed in the introduction, for strictly 2D materials, only $\Omega^{(z)}$ is nonzero, so the anomalous velocity term, $\dot{\mathbf{k}}\times\boldsymbol{\Omega}_{n, \k}$, produces only in-plane Hall-like responses \cite{Ryo2011rotational,nakata2015wiedemann}. In contrast, for few-layer quasi-2D systems, additional in-plane components, $\Omega^{(x)}$ and $\Omega^{(y)}$, allow an out-of-plane anomalous velocity to exist, thus enabling vertical magnon transport.

While the magnetic field enters the semiclassical equations (\eq{eqm1}, (\ref{eqm2})) directly, the effect of a temperature gradient can be incorporated through boundary conditions. To do so and to obtain the contribution of the temperature gradient $(\nb T)$ to the induced magnon-current, we follow the confinement potential approach of \inlinecite{Ryo2011rotational}. In this approach, a sharply rising confining potential is introduced at the sample boundaries to confine the magnons within the system. The potential serves as a mathematical framework to encode the effect of physical boundaries and is eventually integrated out of the final expressions \cite{Ryo2011rotational,Rathor2024spin}. 

The resulting expression for the average vertical magnon current density from the anomalous velocity term in \eq{eqm1}, now dependent on both magnetic field and temperature profiles, is given by (see details in \app{app:vertical_current}):
\begin{align}
\begin{aligned}
\text{j}^{\rm s}_{z}&=\frac{1}{V^2}\sum_{n,\, \k}\,\mu^{\rm s}_{n,\k}\,\int_{V}\,d^3\r\\
&~~~~~~\int_{0}^\infty d \varepsilon \big[\nb\varrho(E_{n,\k}-\bm{\mu}_{n,\k}\cdot \B(\r)-\varepsilon, T(\r))\times \bm{\Omega}_n\big]_z\label{eq:jz}.
\end{aligned}
\end{align}
In the above equation, the index `${\rm s}$' denotes the component of the magnon spin-angular momentum, and the integration over the volume ($V$) of the sample is performed to estimate the average current density. The integration over an additional energy variable $\varepsilon$ appears as a result of integrating out the confinement potential. The application of field and temperature gradients sets up a non-equilibrium steady state in the system leading to the non-equilibrium distribution function $\vrho(E_{n,\k}-\bm{\mu}_{n,\k}\cdot \B(\r)-\varepsilon, \;T(\r))$ for the magnons. A perturbative solution for $\vrho$, in the relaxation time approximation, can be obtained from the  Boltzmann transport equation (BTE): 
\begin{align}
\partial_t\vrho+\dot{\r}\cdot \nb_{\r}\,{\vrho}+\dot{\k}\cdot \nb_{\k}\,{\vrho}=-\frac{({\vrho}-\rho)}{\tau}, 
\end{align}
where $\rho(E_{n,\k}-\bm{\mu}_{n,\k}\cdot \B(\r)-\varepsilon,\; T(\r))$ is the local equilibrium Bose-Einstein distribution, i.e., $\rho(w,T)=n_B(w,T)\equiv[\exp(w/T)-1]^{-1}$ with the temperature $T$ expressed in units of Boltzmann constant k$_B$, and $\tau$ is the magnon relaxation time. The magnon wave packet velocity, $\dot{\r}$, and rate of momentum change, $\dot{\k}$, are given by \eq{eqm1}, \eq{eqm2}, respectively. The steady state ($\partial_t \vrho=0$) solution of BTE, up to first order in gradients, will be 
\begin{align}
\vrho=&\rho-\tau \,(\dot{\r}\cdot \nb_{\r}{\rho}+\dot{\k}\cdot \nb_{\k}\rho)\,+{\cal O}(\nb^2).\label{eq:sol_BTE}
\end{align}
Considering the temperature and magnetic fields to vary linearly along the plane of the sample, we express their spatial profiles as $T(\r)= T_0+(\nabla_\a T) r_\a$ and  $\B(\r)= \B_0+(\nabla_\a \B) r_\a$, respectively, where the index ``$\a$'' sums over the $x,y$ components of the gradients.
The quantities $T_0$ and $\B_0$ are the mean values of the temperature and magnetic field about which their profiles vary.
A Taylor expansion of the RHS term in \eq{eq:sol_BTE} around $T_0$ and $\B_0$ yields a series expansion for $\vrho$ which can then be substituted in \eq{eq:jz} to express the current density in  powers of $\nabla T$ and $\nabla \B$  to give (up to 2nd-order)
\begin{align}
\begin{aligned}
\text{j}^{\rm s}_{z}= &\;\sigma^{\rm s}_{\a}(\nabla_\a T)+\sigma^{\rm s r }_{\a}(\nabla_\a B_{\rm r})+D^{\rm s}_{\a \b}\, (\nabla_\a T)(\nabla_\b T)\\
&~~~~~~+D^{\rm sr}_{\a\b}\, (\nabla_\a T)(\nabla_\b B_{\rm r})+D^{\rm srl}_{\a\b}\, (\nabla_\a B_{\rm r})(\nabla_\b B_{\rm l}),\label{eq:jz_main}
\end{aligned}
\end{align}
where $\sigma_\a^{\rm s}$ and $\sigma_\a^{\rm sr}$ denote the linear conductivities, while $D_{\a\b}^{\rm s}$, $D_{\a\b}^{\rm sr}$ and $D_{\a\b}^{\rm srl}$ represent the nonlinear conductivities induced by temperature and magnetic-field gradients, respectively (see \app{app:vertical_current} for the detailed derivation). For systems having collinear magnetic order with $\k$-independent $\bm{\mu}$ and $\tau$, these conductivities associated with VMT are given by the in-plane BC components $\Omega^{\c=(x,y)}$ as follows
\begin{subequations}
\begin{align}
\sigma^{\rm s}_{\a}=&\;\frac{1}{ V}\sum_{n,\,\k}\mu_n^{\rm s}\,c_1(\rho_0)\,\epsilon_{\a\c}\,\Omega_n^\c\,\label{eq:chi_1}\\
\sigma^{\rm s r }_{\a}=&\;\frac{1}{ V}\sum_{n,\,\k}\mu_n^{\rm s}\,\mu_n^{\rm r}\,c_0(\rho_0)\, \epsilon_{\a\c}\,\Omega_n^\c\label{eq:chi_2}\\
D^{\rm s}_{\a\b}=&-\frac{\tau}{\hbar  T_0 V }\sum_{n,\,\k}\mu_n^{\rm s}\,c_1(\rho_0)\big(\epsilon_{\a\c}\,\frac{\partial (E_n\Omega_n^\c)}{\partial k_\b}+\a \leftrightarrow \b\big)
\label{eq:chi_3}\\
D^{\rm s r l}_{\a\b}=&-\frac{\tau}{ \hbar V}\sum_{n,\,\k}\mu_n^{\rm s}\,\mu_n^{\rm r}\,\mu_n^{\rm l} c_0(\rho_0) \big(\epsilon_{\a\c}\frac{\partial \Omega_n^\c}{\partial k_\b}+\epsilon_{\b\c}\frac{\partial \Omega_n^\c}{\partial k_\a}\big)\label{eq:chi_4}\\
D^{\rm s r }_{\a\b}=& -\frac{\tau}{ \hbar V}\sum_{n,\,\k}\mu_n^{\rm s}\,\mu_n^{\rm r}\,c_1(\rho_0) \big(2\epsilon_{\a\c}\frac{\partial \Omega_n^\c}{\partial k_\b}+\epsilon_{\b\c}\frac{\partial \Omega_n^\c}{\partial k_\a}\big)\label{eq:chi_5}
\end{align}\label{eq:chis}
\end{subequations}
with $\rho_0=n_B(E_{n,\k}, T_0)$, $c_0(\rho_0)=\rho_0,\;c_1(\rho_0)=(1+\rho_0)\ln(1+\rho_0)-\rho_0 \ln(\rho_0)$ and $\epsilon_{\a\b}=-\epsilon_{\b\a}$ is the  Levi-Civita anti-symmetric tensor of rank two. Since we are interested in the transport coefficients in the zero $\B$-field limit, we have set the mean value $\B_0 =0$ in the above equations.
The linear transport responses, $\sigma^{\rm s}_\a$ and $\sigma^{\rm sr}_\a$  are $\tau$ independent and correspond to weighted integrals (or sums) of the hidden components of the symplectic Berry curvature \cite{Peano2018JMP,Peano2016PRX,Shindou2013PRB,Erland2025quantum} (explained in \Sec{Sec:Hidden_BC}) over momenta $\k$. The said responses are similar to the Chern-conductivity for electrons \cite{xiaoBerryPhaseEffects2010}, though not quantized since magnons are bosons. Importantly, the linear responses can only be finite in systems with broken time-reversal symmetry in the single magnon sector and when atleast one layer is non-identical from the others (see \app{app:BdG_BC} for details). In contrast, the non-linear responses $D_{\a\b}^{\rm s}$, $D_{\a\b}^{\rm sr}$ and $D_{\a\b}^{\rm srl}$ are of order $\tau$  and are {given by the} weighted integrals of the hidden components of magnon Berry curvature dipoles and extended magnon BCD \cite{Rathor2024spin,Kondo2022nonlinear,Sodemann2015quantum}
. Accordingly, as mentioned in the introduction, we name the nonlinear coefficients $D_{\a\b}^{\rm s}$, $D_{\a\b}^{\rm s r}$ and $D_{\a\b}^{\rm s r l}$ as hidden magnon BCD and hidden extended magnon BCD, respectively. Unlike the hidden linear coefficients, the hidden non-linear responses, are allowed to be non-zero for  time reversal symmetric layered 2D systems, provided at least one layer is nonidentical.

The expression for vertical magnon current in \eq{eq:jz_main} and the transport coefficients in \Eq{eq:chis} are one of the main results of this paper which we use in \Sec{Sec:models} to compute vertical conductivities of realistic materials. We also note here that temperature and field gradients generate additional in-plane magnon currents that couples to the usual $\Omega^{(z)}$ component of the BC; we have derived the transport coefficients for these responses as well in \app{app:vertical_current}.

\section{Hidden BC in magnon quasi-2D crystals}\label{Sec:Hidden_BC} 
In this section, we identify the origin of in-plane, or hidden, Berry curvature in few-layered quasi-2D magnets. Specifically, our goal will be to express the in-plane BC components -- $\Omega^{(x)}$ and $\Omega^{(y)}$ -- in terms of the magnon excitations of a quasi-2D system, described by the Bloch wavefunctions of the magnon Hamiltonian.

In strictly 2D materials, the out-of-plane crystal momentum $k_z$ is absent and therefore only the $\Omega^{(z)}$ component of BC is non-zero. However, for quasi-2D materials, $k_z$ gets introduced through the dimensional crossover from 3D to 2D, and its effects appear as a limiting case of 3D systems. Consequently, the quasi-2D system can acquire additional components of BC as well as orbital magnetic moments \cite{Hara2020current}. Furthermore, these quantities can also be independently estimated from the intrinsic properties of quasi-2D systems alone, without referring to a 3D parent structure.  For multilayer electronic systems, these quantities have been calculated using a position-like operator representing the inter-layer hoppings \cite{bellissard1995noncommutative}. However, magnon Bloch Hamiltonians of spin systems generally take a bosonic Bogoliubov-de Gennes (BdG) form. Hence, the associated bosonic Berry curvature, often called the symplectic Berry curvature \cite{fang2024quantum,tesfaye2024quantum,Peano2018JMP}, is calculated using paraunitary matrices, in contrast to electronic systems where it is obtained from unitary matrices. Therefore, the position-operator method cannot be directly applied to magnonic, or more generally,  bosonic systems. Instead, as we show below, a generalization of the position operator can be constructed for magnon BdG systems leading to the identification of a pseudo-position operator that enables us to quantify the hidden BC of magnon modes.

To get to the Bloch Hamiltonian and the magnon modes of a spin system with a stable collinear magnetic order, we employ linear spin wave theory (LSWT). In this approach, the spin operators ($\S_i$) of a Hamiltonian $H=\sum_{ij} \S_i^T \textbf{J}_{ij}\S_{j}$, describing spin-spin interactions of strength $\textbf{J}_{ij}$ between lattice sites $i$, $j$, are first mapped to site-local bosonic excitations ($a_i$, $a_i^\dagger$)  using the Holstein-Primakoff (HP) transformations
\[
\S_i^- = (2S - a_i^\dagger a_i)^{\frac{1}{2}} a_i^\dagger,~
\S_i^+ = a_i (2S - a_i^\dagger a_i)^{\frac{1}{2}},~
\S_i^z = S - a_i^\dagger a_i
\]
where $\S^\pm_i\equiv \S^x_i\pm i\,\S^y_i $,  and $S$ is the ground state expectation value of the local spin order. Next, expanding $H$ in powers of $1/S$ (see \app{app:buckled_HC_AF}), the linear term in $S$ yields a real-space lattice magnon Hamiltonian, while $O(1/S)$ corrections are neglected assuming $S$ to be large. Performing a Fourier transform of the real space Hamiltonian, produces the magnon Bloch Hamiltonian, which takes the bosonic-BdG form \cite{Matsumoto2014Thermal,Kondo2020Non}

\begin{align}
H=&\sum_\k \psi_\k^\dagger\H_\k \psi_\k=\sum_\k \begin{bmatrix}\bm{a}^\dagger_\k~ \bm{a}_{\tminus\k}\end{bmatrix}\!\!\begin{bmatrix}\! h_\k &\Delta_\k\\ \Delta^*_{\tminus\k} &h^*_{\tminus\k}
    \end{bmatrix}\!\!\begin{bmatrix}
        \bm{a}_\k\\ \bm{a}^\dagger_{\tminus\k}
    \end{bmatrix}\label{eq:BdG_Ham}
\end{align}
where $h_\k=h^\dagger_\k$ and $\Delta_\k=\Delta^T_{\tminus\k}$  are $N\times N$ matrices, with $N$ being the number of sub-lattice sites in a unit-cell of the system.  
The symbol $\bm{a}_\k\equiv (a_{1\k},a_{2\k},\cdots ,a_{N\k})$ represents the set of bosonic operators $a_{\alpha \k}$ residing on the sub lattice $\alpha$ within a unit-cell and obtained from the Fourier transform of real space operators $a_i$.
 The sub-lattice operators $a_{\alpha\k}$, $a^\dagger_{\beta\k}$ follow bosonic commutations --
$[a_{\alpha\k},a^\dagger_{\beta\k'}]=\delta_{\alpha\beta}\delta_{\k\,\k'}$, implying the Nambu spinor \cite{Ohashi2020Generalized} $\psi_\k\equiv [\bm{a}_\k,~\bm{a}^\dagger_{\tminus\k}]^T$ satisfies the commutation relations $[\psi_\k,\psi^\dagger_{\k'}]=(\Sigma_z)_{\alpha\beta}\delta_{\k\,\k'}$. The matrix $\Sigma_z\equiv\sigma_z\otimes \bm{1}_N$ comprises the Pauli matrix, $\sigma_z=\text{diag}(1,-1)$, acting on the BdG (particle-hole) sector \cite{Kondo2020Non}, and $\bm{1}_N$ denotes the identity matrix  belonging to the sub-lattice space.

Unlike the electron-BdG system, an arbitrary unitary transformation of $\psi_{\k}$ does not preserve the bosonic commutations discussed above. Consequently, the coefficient matrix $\H_\k$ (in \eq{eq:BdG_Ham}) cannot be in general diagonalized using unitary matrices to yield the independent magnon modes. Instead, transformations $\psi'_\k=\P_\k \psi_\k$, generated by a para- (or pseudo-) unitary matrix $\P_\k$ satisfying 
\begin{align}
\P_\k^\dagger \Sigma_z\P_\k =\Sigma_z,\label{eq:para_unitary}
\end{align}
preserve the commutation relations, i.e. $[\psi'_\k,{\psi'}^\dagger_\k]=\Sigma_z$, and hence can be used to find the diagonal modes. Therefore, one needs to solve the eigenvalue problem \cite{Matsumoto2014Thermal,Kondo2020Non}
\begin{equation}
\P_\k^\dagger \H_\k\P_\k =\text{diag}(E_{1,\k},\cdots,E_{N,\k},E_{1,\tminus\k},\cdots,E_{N,\tminus \k}), 
\end{equation}
under the constraint in \eq{eq:para_unitary}, to find the magnon band energies $E_{n,\pm \k}$ and the paraunitary matrices $\P_\k$ containing the magnon eigenmodes.

Since the mathematical condition in \eq{eq:para_unitary} is analogous to the constraints satisfied by elements of the symplectic group $\mathrm{Sp}(2N,\mathbb{R})$, quantities involving paraunitary matrices are often described using the qualifier symplectic. Consequently, the Berry curvatures for the bosonic BdG system, given by (see \app{app:BdG_BC}, \cite{Matsumoto2014Thermal,Shindou2013PRB})
\begin{align}
\Omega^{\a}_{n\k}\equiv \epsilon_{\a\b\c}\,\Im\Big[\Sigma_z\frac{\partial \para^\dagger_\k}{\partial {k_\b}}\Sigma_z\frac{\partial \para_\k}{\partial {k_\c}}\Big]_{nn}\label{eq:BCmunu}
\end{align} 
are referred to as symplectic Berry curvatures and depend on both $\Sigma_z$ and the derivatives of the paraunitary matrix $\para_\k$.

As seen from the expression above, the in-plane components $\Omega^{\a =(x,y)}$ require derivatives of $\para_\k$ with respect to $k_z$, which are a priori zero in quasi-2D systems since $k_z$, while present, does not explicitly appear in the bare quasi-2D Hamiltonian. To illustrate this point, we consider a quasi-2D spin system consisting of $L$ layers  parallel to the $xy$-plane  whose magnons are described by a prototypical BdG Hamiltonian
 \begin{equation}
\H_\text{quasi-2D}(\k)= \!\left[\begin{array}{ll}
\big[~\,\mathbf{h}_{ll'}(\k)~\,\big]_{\scriptscriptstyle L\times L} & \big[\,\mathbf{\Delta}_{ll'}(\k)\big]_{\scriptscriptstyle L\times L}\\[8pt]
\big[\mathbf{\Delta}^*_{ll'}(\tminus\,\k)\big{]}_{\scriptscriptstyle L\times L} & \big[\mathbf{h}^*_{ll'}(\tminus\,\k)\big]_{\scriptscriptstyle L\times L} 
\end{array}\right]\label{eq:Ham_quasi2D}
\end{equation}
where the elements, $\mathbf{h}_{ll'}$, $\mathbf{\Delta}_{ll'}$,  of the $L\times L$ sub-matrices can themselves be  matrices having dimensions consistent with the subunit-cells across layers. The diagonal blocks $\mathbf{h}_{ll}$, $\mathbf{\Delta}_{ll}$ represent couplings within the $l$-th layer, and 
$\mathbf{h}_{l\,l'\neq l}$, $\mathbf{\Delta}_{l\,l'\neq l}$, encode interlayer couplings. 
More importantly, since lattice-translation symmetry is absent in the $\hat{\textbf{z}}$-direction, the Bloch momenta $\k=(k_x,k_y)$ that enter into $\H_\text{quasi-2D}$ comprises only the in-plane components $k_x$ and $k_y$.

 To incorporate the effect of $k_z$, we will generalize the dimensional crossover idea highlighted in the introduction and in the beginning of this section. To this end, we build a 3D system in the vertical direction using weakly coupled copies of the quasi-2D Hamiltonian in \eq{eq:Ham_quasi2D} as the motif and arrive at the $k_z$-dependent Hamiltonian
\begin{equation}
\!\!\!\H_\text{3D}(\k,k_z)=\!\!\begin{bmatrix}
\big[~~\mathbf{\tilde{h}}_{ll'}(\k,k_z)~~\big]_{\scriptscriptstyle L \times L} \!& \big[~\mathbf{\tilde{\Delta}}_{ll'}(\k,k_z)~\big]_{\scriptscriptstyle L\times L}\\[10pt]
\big[\mathbf{\tilde{\Delta}}^*_{ll'}(\tminus\k,\tminus k_z)\big]_{\scriptscriptstyle L\times L} \!& \big[\mathbf{\tilde{h}}^*_{ll'}(\tminus\k,\tminus k_z)\big]_{\scriptscriptstyle L\times L} 
\end{bmatrix}\label{eq:Ham_extended3D}
\end{equation}
with,
\begin{align}
\mathbf{\tilde{h}}_{ll'}(\k,k_z)=& (\mathbf{h}_{ll'}(\k)+\mathbf{h}'_{ll'}e^{i\delta k_z}+{\mathbf{h}'}^\dagger_{l'l}e^{-i\delta k_z})e^{i(\xi_{l'}-\xi_l) k_z}\notag\\
\mathbf{\tilde{\Delta}}_{ll'}(\k,k_z)=&(\mathbf{\Delta}_{ll'}(\k)+\mathbf{\Delta}'_{ll'}e^{i\delta k_z}+{\mathbf{\Delta}'}^T_{l'l}e^{-i\delta k_z})e^{i(\xi_{l'}-\xi_l) k_z},  \notag 
\end{align}
where $\xi_{l=1,2,\cdots,L}$ are the $z$-positions of the layers relative to some reference point within each copy and $\delta$ is the $z$-direction separation between  equivalent lattice points of any two adjacent copies. The additional matrices $\mathbf{h}'_{ll'}$ and $\mathbf{\Delta}'_{ll'}$ represent the coupling between adjacent copies connecting layer $l$ of one copy to layer $l'$ of another.

Observing that, in the limit of zero coupling between copies, $\H_{3D}$ relates to $\H_\text{quasi-2D}$ through
\begin{align*}
\H_{3D}(\k,k_z)=\para_{k_z}\H_\text{quasi-2D}(\k) \para^\dagger_{k_z}
\end{align*}
via the $k_z$-dependent matrix $\para_{k_z}=I_2\otimes\text{diag}( e^{i \xi_1k_z},\cdots ,e^{i \xi_Lk_z})\notag$, we realize that the paraunitary transformation $\para_{3D}$ diagonalizing $\H_{3D}$ is given by
\begin{align}
\para_{3D}(\k,k_z)=\para_{k_z} \para_\k,\label{eq:para_product}
\end{align}
where $\para_\k$ is the paraunitary matrix that diagonalizes $\H_\text{quasi-2D}$ (see \app{app:BdG_BC} for details). Following this insight and using \eq{eq:BCmunu}, we write the in-plane components $(\a = x,y)$ for the bulk-3D system in terms of $\para_{k_z}$ and $\para_\k$, as follows:
\begin{align}
{\Omega}_{n\k}^{\a}=&\,2\epsilon_{\a\b}\,\Im\Big[\Sigma_z \frac{\partial \para_{3D}^\dagger(\k,k_z)}{\partial k_z} \Sigma_z\frac{\partial \para_{3D}(\k,k_z)}{\partial k_\b}\Big]_{nn}\notag\\
=&\,2\epsilon_{\a\b}\,\Im\Big[\Sigma_z\para^\dagger_\k\Big( \frac{\partial \para^\dagger_{k_z}}{\partial k_z} \Sigma_z \para_{k_z}\Big)\frac{\partial \para_{\k}}{\partial k_\b}\Big]_{nn}.\notag \end{align}
Evaluating the $k_z$-derivative and simplifying (see \app{app:BdG_BC}), we can finally express the in-plane symplectic BC components as
\begin{align}
{\Omega}_{n\k}^{\a}=2\epsilon_{\a\b}\,\Re\Big[ \Sigma_z\para^\dagger_\k\,\Z\frac{\partial \para_\k}{\partial k_\b}\Big]_{nn},\label{eq:hidden_BC}
\end{align}
written entirely in terms of quantities intrinsic to the quasi-2D system. In the process, we identify the operator
\begin{equation} \Z=\Sigma_z\otimes\text{diag}( \xi_1,\xi_2,\cdots,\xi_L),\label{eq:pseudo_Z}
\end{equation}
which we define as the \emph{pseudo-position operator,} since it acts non-trivially on the BdG sector and contains the inter-layer distances of the quasi-2D system.
Moreover, since $\Omega^{(x,y)}$ depend only on the properties of the quasi-2D Hamiltonian (\eq{eq:Ham_quasi2D}, \eq{eq:hidden_BC}), these in-plane components are naturally expected to remain finite even when our 3D construction (\eq{eq:Ham_extended3D}) is reduced to a single copy of the original quasi-2D structure (\eq{eq:Ham_quasi2D}) consisting of a few atomic layers. Thus, the in-plane symplectic-BC components, $\Omega^{(x,y)}$, emerge as hidden magnon Berry curvatures in the quasi-2D magnet. For practical purposes \eq{eq:hidden_BC} can be recast into the form
\begin{align}
\Omega_{n\k}^{\a}=\epsilon_{\a\b}\,\frac{\partial}{\partial k_\b}\big[\Sigma_{z}\para^\dagger \Z\,\para\big]_{nn}\label{eq:hidden_BC_formula}
\end{align} 
expressed explicitly in terms of the matrix elements of the respective operators. The expression for the hidden magnon BC above and in \eq{eq:hidden_BC} differs from the electronic hidden BC in two major ways: first, the transformations $\para_{\k}$ are paraunitary rather than unitary; and second, the pseudo-position operator $\Z$ appears in place of a standard position operator.

To summarize, \Eq{eq:hidden_BC} which defines the hidden magnon BC and \Eq{eq:pseudo_Z} introducing the pseudo-position operator are the main findings of this section and another key result of our work. These equations, along with \Eq{eq:jz_main} and \Eq{eq:chis} of \Sec{Sec:VMT}, form the complete set needed to compute VMT in collinear quasi-2D magnets.

\section{Vertical transport in realistic models}\label{Sec:models}
In this section, we  study two realistic spin models -- the buckled honeycomb lattice and bilayer-CrX$_3$ (X = Cl, Br, I). We find both these systems show non-vanishing hidden magnon Berry curvatures (HMBC) and we estimate  the  vertical magnon transport due to magnetic field gradient $\nabla \B$ and temperature gradient $\nabla T$ as derived in \Sec{Sec:VMT}. To keep the presentation simple, we focus on bilayer spin systems, though the discussion can be easily extended to  multilayer cases. Additionally, going by experimental reports we consider the models to have collinear out-of-plane magnetic order so that only the $\hat{\textbf{z}}$-component of magnetic moment for the magnons is non-zero, i.e., $ \bm{\mu}_n=(0,0,\mu_n)$. In this setting, the following components of the response tensors derived in \eq{eq:chis} will produce finite contributions to VMT:
\begin{enumerate}
\item \textbf{Linear response} components originating from total hidden Berry curvature contributions 
\renewcommand*{\arraystretch}{1.2}
\begin{equation}
 \begin{array}{cc}
 \sigma_x^z =\frac{1}{ V}\sum\limits_{n,\,\k}\mu_n c_1(\rho_0)\Omega_n^y,&\sigma_y^z = -\frac{1}{ V}\sum\limits_{n,\,\k}\mu_n c_1(\rho_0)\Omega_n^x, \\
 \sigma_x^{zz} =\frac{1}{ V}\sum\limits_{n,\,\k}\mu^2_n c_0(\rho_0)\Omega_n^y,&\sigma_y^{zz} = -\frac{1}{ V}\sum\limits_{n,\,\k}\mu^2_n c_0(\rho_0)\Omega_n^x 
 \end{array}  \label{eq:L_VMT}
\end{equation}
\item \textbf{Non-linear response} components arising from hidden Berry curvature dipole and hidden extended Berry curvature dipole 
\renewcommand*{\arraystretch}{1.25}
\begin{equation}
 \begin{array}{c}
D^{z}_{\a\b}=-\frac{\tau}{\hbar^2 V T_0}\sum\limits_{n,\,\k}\mu_nc_1(\rho_0)\;\chi^{\scriptscriptstyle(1)}_{\a\b}, \\
D^{zz}_{\a\b}~~=~~-\frac{\tau}{\hbar^2 V}\sum\limits_{n,\,\k}\mu_n^2c_1(\rho_0)\;\chi^{\scriptscriptstyle(2)}_{\a\b}, \\
D^{zzz}_{\a\b}~=~~-\frac{\tau}{\hbar^2 V}\sum\limits_{n,\,\k}\mu_n^3c_0(\rho_0)\;\chi^{\scriptscriptstyle(3)}_{\a\b}, 
 \end{array}   \label{eq:NL_VMT}
\end{equation}
where $\chi^{(1,2,3)}_{\a\b}$ ($\a,\b = \{x,y\}$) are defined as
\renewcommand*{\arraystretch}{1.9}
\begin{center}
\begin{equation}
 \begin{array}{|c|ccc|c|c|}\hline
  & xx&xy,yx&yy\\\hline
\chi^{\scriptscriptstyle(1)}_{\a\b}&\frac{\partial (E_n\Omega_n^y)}{\partial k_x}&\frac{1}2\big(\frac{\partial (E_n\Omega_n^y)}{\partial k_y}-\frac{\partial (E_n\Omega_n^x)}{\partial k_x}\big) &-\frac{\partial (E_n\Omega_n^x)}{\partial k_y}  \\\hline
\chi^{\scriptscriptstyle(2)}_{\a\b}&\frac{3}{2}\frac{\partial \Omega_n^y}{\partial k_x}&\frac{\partial \Omega_n^y}{\partial k_y}-\frac{1}{2}\frac{\partial \Omega_n^x}{\partial k_x}&-\frac{3}{2}\frac{\partial \Omega_n^x}{\partial k_y} \\\hline
\chi^{\scriptscriptstyle(3)}_{\a\b}&\frac{\partial \Omega_n^y}{\partial k_x}&\frac{1}{2}\,\big(\frac{\partial \Omega_n^y}{\partial k_y}-\frac{\partial \Omega_n^x}{\partial k_x}\big)&-\frac{\partial \Omega_n^x}{\partial k_y}  \notag\\\hline
 \end{array}  
\end{equation}
 \end{center}
\end{enumerate}
and the definitions for rest of the symbols, $\rho_0$, $T_0$ etc., are given near \eq{eq:chis}.

To ensure a finite response from the hidden magnon Berry curvatures, the layers of the quasi-2D magnets must be non-identical (see \app{app:BdG_BC}). From an experimental perspective, layer asymmetry could be introduced in several ways, such as, by tuning interactions of the top or bottom layers with the substrate, or by applying dissimilar strain fields to the layers. Particularly for layered antiferromagnets having distinct ordering directions on individual layers, the layer symmetry is broken naturally by magnetic order, without the need to break the  symmetry explicitly.

In the bilayered systems studied in this section, we model such layer-symmetry breaking mechanisms, in either or both of the following ways: one, by setting spin Hamiltonian parameters, such as Heisenberg couplings, easy-axis anisotropy, etc., to be distinct for the two layers; and two, by choosing opposite ordering directions across the two layers for anti-ferromagnets.
We then track the behavior of the linear and non-linear responses (see \eq{eq:L_VMT} and \eq{eq:NL_VMT}) as functions of experimentally tunable parameters, such as temperature, interlayer couplings, etc., and present observations on their salient features.

\begin{figure*}[t]
\includegraphics[width=1\textwidth]{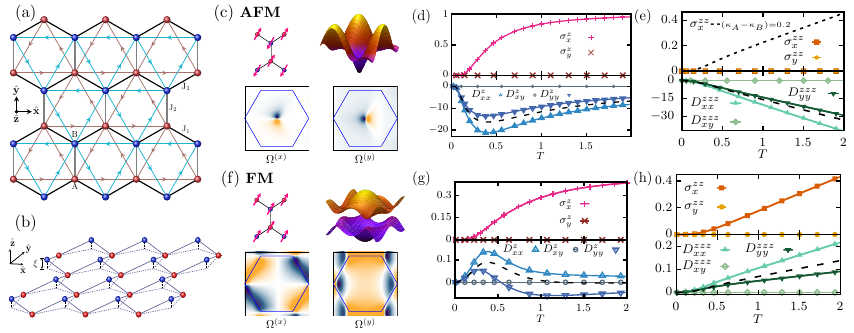}
\vspace*{-.25cm}
\caption{\textbf{Buckled honeycomb lattice} (a) Top view of the lattice showing Heisenberg interactions, with coupling strength $\J_1$ and $\J_2$, connecting sublattice sites A with B, and Dzyaloshinskii Moriya (DM) interactions (directions shown by arrows) coupling sites of the same sublattice  with strength $D_A$, $D_B$ for $A$, $B$ sublattices, respectively. Panel (b) shows the side view of the lattice, $\xi$ is the relative vertical shift between sublattices A and B. Panel (c) shows the spin orientations,  typical magnon-bands and the two hidden (in-plane) components of magnon Berry-curvature (BC), $\Omega^{(x)}$ and $\Omega^{(y)}$, for the lowest energy magnon-mode  in the AFM ordered phase. Panel (f) shows the same plots as panel (c) for the FM order. The magnon-BC components within the first BZ (hexagon) are plotted in units of $\xi$ with yellow (blue) regions indicating positive (negative) values. The VMT responses arising from temperature gradient ($\nabla T$) -- $\sigma^z_{x},\sigma^z_{y},D^z_{xx},D^z_{xy}$, $ D^z_{yy}$, and those arising from  magnetic field gradient ($\nabla \B$) -- $\sigma^{zz}_{x},\sigma^{zz}_{y},D^{zzz}_{xx},D^{zzz}_{xy}$, $ D^{zzz}_{yy}$, for the AFM case are plotted in panels (d) and (e), respectively, using parameters   $\J_1=1.0,\J_2=0.7, D_A=-D_B=0.35,\kappa_A=\kappa_B=0.01$. The corresponding VMT responses for the FM case are shown in (g) and (h), respectively, with  $\J_1=1.0,\J_2=0.7, D_A=D_B=0.1, \kappa_A=0.1,\kappa_B=0.01$.}\label{Fig:buckled_HC}
\end{figure*}

\subsection{Vertical magnon transport in buckled honeycomb lattice (BHC)}\label{Sec:Buckled_HC}

As a first example, we study VMT in a simple yet popular spin model of a two-band topological magnon insulator on the honeycomb lattice \cite{Fransson2016magnon}. Having only two magnon bands allows analytical evaluation of the band dispersions and Berry curvatures, making it ideal for conveying the proof of concept. The model (see \subfig{Fig:buckled_HC}{a}) includes nearest-neighbor Heisenberg and next-nearest-neighbor Dzyaloshinskii-Moriya (DM) interactions and exhibits topological edge states analogous to those in the Haldane model \cite{owerre2016first}. The DM interactions break time-reversal symmetry in the magnon Hamiltonian, allowing the magnon bands to acquire finite Chern numbers from the out-of-plane component $\Omega^{(z)}$ of the Berry curvature. Consequently, the conventional in-plane thermal Hall effect arises from the finite $\Omega^{(z)}$ component, while the hidden magnon BC components remain zero for a perfectly flat honeycomb lattice.
 
  If the two sublattices $A$ and $B$ of the honeycomb lattice develop out-of-plane displacements due to buckling (see \subfig{Fig:buckled_HC}{b}), hidden Berry curvatures can become finite and contribute to vertical transport. The buckled-honeycomb lattice (BHC) has been studied extensively in electronic materials \cite{Huang2014Ferromagnetism,Yang2015Buckled} and in spin systems such as Ba$_2$NiTeO$_6$ \cite{Asai2017Spin}.
This system can be interpreted as two flat triangular-lattice layers, vertically separated by a distance $\xi$ (see \subfig{Fig:buckled_HC}{b}), where the first (second) layer is formed by $A$ ($B$) sublattice sites.

The spin Hamiltonian for the magnon insulator on the BHC lattice is given by \cite{owerre2016first}
\begin{align}
H=&\sum_{\scriptscriptstyle\nn{i}{j}}{\rm J}_{ij}\,\S_{i}^{\scriptscriptstyle (A)}\cdot\S_j^{\scriptscriptstyle (B)}+\sum_{\scriptscriptstyle l \in A, B}D_l\sum_{\scriptscriptstyle\langle\langle{i},{j}\rangle\rangle}\nu_{ij}(\S^{(l)}_{i}\!\times\S^{(l)}_j)^z\notag\\
&~~~~~~~~~~~~~~~-\kappa_A\sum_{\scriptscriptstyle i\in A}\big(\S^{z}_i\big)^2-\kappa_B\sum_{\scriptscriptstyle i\in B}\big(\S^{z}_i\big)^2,\label{eq:BHC_spin_Ham}
\end{align} 
where the vertical displacements of the buckled layers are introduced through the position coordinates of the sites $i$, $j$.
The first term in the above Hamiltonian represents nearest-neighbor Heisenberg interactions with coupling strengths, $\J_1$ for slanted bonds and $\J_2$ for vertical bonds, see \subfig{Fig:buckled_HC}{a}. Positive (negative) $\J_{1,2}$ will stabilize anti-ferromagnetic (ferromagnetic) order on the two sub-lattices $A$ and $B$. The second term {incorporates} next nearest neighbor DM interactions with strengths $D_A$ and $D_B$ for the two sub-lattices $A$ and $B$, respectively with their directions (accounted by $\nu_{ij}=\pm 1$) shown in \subfig{Fig:buckled_HC}{a}. The last two terms represent easy axis anisotropies $\kappa_A$ and $\kappa_B$ for the two sub-lattices and stabilizes the magnetic ordering of the spins aligned along the out-of-plane direction.\subsubsection{Anti-ferromagnet}\label{Sec:buckled_HC_AF}
In the anti-ferromagnetic (AFM) phase ($\J_{1,2}>0$), the spins on the sublattices $A$ and $B$,  are ordered antiparallel to each other along the $z$-axis. As discussed in \Sec{Sec:Hidden_BC} (and detailed in \app{app:buckled_HC_AF}), the magnon Hamiltonian is obtained in two steps -- First, applying the HP transformations
\begin{align*}
\begin{split}
   \S^{\scriptscriptstyle (A)-}_i=& \;(2S-a^\dagger_i a_i)^{\frac{1}{2}}a^\dagger_i,\\
   \S^{\scriptscriptstyle (A)+}_i=&\;a_i(2S-a^\dagger_i a_i)^{\frac{1}{2}},\\
   \S^{\scriptscriptstyle (A)z}_i=&\;S-a^\dagger_i a_i
\end{split}
~~\Bigg |~~
\begin{split}   
   \S^{\scriptscriptstyle (B)-}_i=& \;b_i(2S-b^\dagger_i b_i)^{\frac{1}{2}},\\
   \S^{\scriptscriptstyle (B)+}_i=&\;(2S-b^\dagger_i b_i)^{\frac{1}{2}}b^\dagger_i,\\
   \S^{\scriptscriptstyle (B)z}_i=&-S+b^\dagger_i b_i
\end{split}     
\end{align*}
to represent the spins $S_i$ in terms of local bosonic operators $a_i$, $b_i$, residing on $A$, $B$ sublattices, to get to the Fourier modes
\begin{equation}
\bigg[\begin{array}{cc}
a_\k  \\b_\k  \end{array}\bigg]
=\sum_{\R_i}e^{-i\k\cdot \R_i} \bigg[\begin{array}{cc}e^{-i\k\cdot \r_A} a_i  \\[1pt]e^{-i\k\cdot \r_B} b_i \end{array}\bigg],\label{eq:Fourier_transform}
\end{equation}
where $\k$ is the Bloch-momentum, $\R_i$ iterate over Bravais-lattice vectors and $\r_{A,B}$ are the relative positions of the spins within the unit-cell. Next, substituting the above modes into the spin Hamiltonian (\eq{eq:BHC_spin_Ham}) and retaining the bilinear terms in the bosonic operators  yields the magnon-BdG Hamiltonian:
\begin{equation}
    H_\text{magnon}=\frac{1}{2}\sum_\k \bm{\psi}^\dagger_\k\bigg[\begin{array}{cc}
        h_\k &\Delta_\k\\
        \Delta^*_{\tminus\k}&h^*_{\tminus \k}
    \end{array}\bigg] \bm{\psi}_\k\label{eq:Ham_BHC_AF}
\end{equation}
with $\bm{\psi}^\dagger_\k=[a^\dagger_\k\;b^\dagger_\k \;a_{\tminus\k}\;b_{\tminus\k}]$,  
\begin{equation}
  h_\k=\bigg[\begin{array}{cc}
d_{A\k}&0\\0&d_{B\k}
\end{array}\bigg], ~~~~   \Delta_\k=\bigg[ 
\begin{array}{cc}
 0&\gamma_\k\\ \gamma_{\tminus\k}&0
\end{array}\bigg],  \label{eq:h_and_Delta}
\end{equation}
and
\begin{align*}
\gamma_\k=&\;2\J_1S\,\cos(k_x/2)e^{i\frac{k_y}{2\sqrt{3}}}+\J_2 Se^{-i\frac{k_y}{\sqrt{3}}},\\    
d_{A,B\k}=&\;(2\J_1+\J_2+2\kappa_{A,B})S\\
&~~+ 2 D_{A,B}S[\sin k_x-2\sin(k_x/2)\cos\big(\sqrt{3}k_y/{2}\big)].
\end{align*}
When $D_A=D_B$, time-reversal symmetry is present in the magnon sector and is broken otherwise (see \app{app:buckled_HC_AF} for explanation).  Additionally, for $\J_2=\J_1$ the model has a three fold rotational symmetry C$_3$ around the $\textbf{z}$-axis. The two independent magnon modes are obtained by diagonalizing the Hamiltonian in \eq{eq:Ham_BHC_AF} using paraunitary transformations (see \Sec{Sec:Hidden_BC}, \app{app:buckled_HC_AF}). The resultant band dispersions for the magnon modes are
\begin{equation}
 E_{\pm,\k}=\pm\frac{(d_{A\k}\rminus\,d_{B\tminus\k})}{2}+\frac{1}{2}\sqrt{(d_{A\k}+d_{B\tminus\k})^2\rminus 4|\gamma_\k|^2},\label{eq:band_dispersion_BHC_AF}
\end{equation}
and their respective magnetic moments are given by
\begin{equation}
    \bm{\mu}_\pm=\pm \,g \mu_B S\,\hat{\textbf{z}},
\end{equation}
where $g$ is the Land{\' e}-$g$-factor and $\mu_B$ is the Bohr-magneton. 
Considering the displacement between the sublattices to be $\xi$ and the definition given in \eq{eq:pseudo_Z},  the
 pseudo-position operator for this model is written as $\Z=\Sigma_z\otimes \text{diag}(\frac{\xi}{2},-\frac{\xi}{2})$. Hence, using the expression in \eq{eq:hidden_BC_formula} and the said $\Z$ operator, we evaluate the hidden BC for the two magnon bands of the BHC lattice to be
\begin{equation}
{\Omega}^{\a}_{\pm, \k}=\frac{\epsilon_{\a\b}\xi}{2}\frac{\partial}{\partial k_\b}\frac{(d_{A\k}+d_{B\tminus\k})}{[(d_{A\k}+d_{B\tminus\k})^2-4|\gamma_\k|^2]^{1/2}}.\label{eq:BC_BHC_AF}
\end{equation}
The opposite  directions of the magnetic order on the two sublattices break layer symmetry, thus ensuring non-zero HMBC for all finite couplings. We plot the BZ distributions for the hidden BC components $\Omega^{(x,y)}$, evaluated using \eq{eq:BC_BHC_AF}, in \subfigs{Fig:buckled_HC}{c}{d}, respectively. The $\Omega^{(x)}$($\Omega^{(y)}$) distribution is antisymmetric (symmetric) about the $k_y=0$ line and is concentrated near the energy-band (\subfig{Fig:buckled_HC}{c}(inset)) bottom corresponding to the $\k=\textbf{0}$ point.

The vertical magnon transport coefficients -- $\sigma^z_x,\sigma^z_y$, $D^z_{xx},D^z_{xy},D^z_{yy}$ generated purely due to the temperature gradient $(\nabla T)$ are plotted in (\subfig{Fig:buckled_HC}{d}), and the coefficients -- $\sigma^{zz}_x,\sigma^{zz}_y$, $D^{zzz}_{xx},D^{zzz}_{xy},D^{zzz}_{yy}$ arising only from the field gradient ($\nabla \B$) are shown in (\subfig{Fig:buckled_HC}{e}). The overall sign of the responses depends on the ordering directions of the spins; reversing the direction of all the spins will change the sign of the responses as well. The linear transport coefficients --$\sigma^{z}_{x,y}$, $\sigma^{zz}_{x,y}$-- are expressed in units of $g\mu_BS$ and $(g\mu_BS)^2$, respectively. The nonlinear coefficients -- $D^{z}_{xx/xy/yy}$,~$D^{zzz}_{xx/xy/yy}$ -- are shown in units of $\tau g\mu_BS/\hbar$ and $\tau(g\mu_B S)^3/\hbar$, respectively. The parameter values used to generate the plots are provided in the caption of \fig{Fig:buckled_HC}. The responses $\sigma^z_y$, $D^z_{xy}$ (for $\nabla T$) and $\sigma^{zz}_y$, $D^{zzz}_{xy}$ (for  $\nabla \B$) are identically zero for all parameters due to an underlying symmetry of the magnon Hamiltonian (\eq{eq:Ham_BHC_AF}) under the transformation $k_y\rightarrow -k_y$. Further, when $\J_1=\J_2$ the system has C$_3$ symmetry making the linear coefficients $\sigma^z_{x,y}$, $\sigma^{zz}_{x,y}$ vanish and constrains the non-linear coefficients to satisfy $D^{z}_{xx}=D^{z}_{yy}$ (dashed line in \subfig{Fig:buckled_HC}{d}) and $D^{zzz}_{xx}=D^{zzz}_{yy}$ (dashed line in \subfig{Fig:buckled_HC}{e}). The linear coefficients also vanish when time-reversal symmetry is restored by setting $D_A=D_B$ (not shown). For $T\rightarrow 0$, all coefficients for both temperature and field gradients approach zero as magnon density vanishes at low temperatures. The $\nabla T$ coefficients also saturate to constant values at higher temperatures (\subfig{Fig:buckled_HC}{d}).

The linear coefficients for $\nabla \B$ are zero when all parameters $\J_{1,2},D_{A,B},\kappa_{A,B}$ are same across layers (see \fig{Fig:buckled_HC} top), consistent with the absence of net magnetization in anti-ferromagnets, which prevents first-order coupling to $\nabla \B$. Making $\kappa_A$ different from $\kappa_B$ can generate a finite value for the linear coefficient for $\sigma^{zz}_x$ (dashed line in \subfig{Fig:buckled_HC}{e}). Both the linear coefficients (when finite) and the non-linear coefficients related to $\nabla \B$ vary linearly at higher temperature values. An overall comparison between the temperature and field coefficients reveal the non-linear coefficients for $\nabla T$ to vary non-monotonically with $T$ (\subfig{Fig:buckled_HC}{d}) which is qualitatively different from the monotonic $T$-dependence of the non-linear coefficients (\subfig{Fig:buckled_HC}{e}) for $\nabla \B$. The coefficients for the cross terms in $\nabla T$, $\nabla \B$ (see \eq{eq:NL_VMT}) are discussed in \app{app:models}.
\subsubsection{Ferromagnet}\label{Sec:buckled_HC_F}
The ferromagnetic (FM) state in the BHC lattice is stabilized when $\J_{1,2}<0$ (see \eq{eq:BHC_spin_Ham}), and the spins residing on sublattice $A$ and $B$ both align in the same direction along the $z$-axis. Performing the HP transformations 
\begin{align}
\begin{split}
  \S^{\scriptscriptstyle (A)-}_i=& \;(2S-a^\dagger_i a_i)^{\frac{1}{2}}a^\dagger_i,\\
   \S^{\scriptscriptstyle (A)+}_i=&\;a_i(2S-a^\dagger_i a_i)^{\frac{1}{2}},\\
   \S^{\scriptscriptstyle (A)z}_i=&\;S-a^\dagger_i a_i,
\end{split}
\Bigg|
\begin{split}   
   \S^{\scriptscriptstyle (B)-}_i=& \;(2S-b^\dagger_i b_i)^{\frac{1}{2}}b^\dagger_i,\\
   \S^{\scriptscriptstyle (B)+}_i=&\;b_i(2S-b^\dagger_i b_i)^{\frac{1}{2}},\\
   \S^{\scriptscriptstyle (B)z}_i=&\;S-b^\dagger_i b_i
\end{split} \label{eq:BHC_FM_HP}    
\end{align}
and retaining terms bilinear in the Fourier modes (see \eq{eq:Fourier_transform})
results in the magnon Hamiltonian 
\begin{equation}
h_\k =\bigg[\begin{array}{cc}
d_{A\k}& \rminus\gamma_\k\\
\rminus\gamma_\k^*&~~d_{B\k}
\end{array}\bigg],~~~~~~\Delta_\k=\bm{0}_{2\times 2}\label{eq:hk_F}
\end{equation}
where $d_{A\k}$, $d_{B\k}$ and $\gamma_\k$ are the same as defined for the antiferromagnetic case (near \eq{eq:h_and_Delta}). Since, the $\Delta_\k$ matrix is always zero for pure ferromagnets, the magnon Hamiltonian for such magnets can be further simplified to 
\begin{align}
H_\text{magnon}=\sum\limits_{\k} \bm{\phi}^\dagger_\k h_\k\,\bm{\phi}_{\k}, ~~~~\text{with}~~~\bm{\phi}_{\k}^\dagger=[a^\dagger_\k, b^\dagger_\k]\label{eq:Ham_BHC},
\end{align}
resembling  a number conserving electronic lattice Hamiltonian and can be described by only two physical bands without their BdG copies.  Solving the above Hamiltonian, we find the energies of the two magnon bands $E_{+,\k},E_{-,\k}$ to be
\begin{equation}
    E_{\pm\k}=\frac{1}{2}(d_{A\k}+d_{B\k})\pm\frac{1}{2}\sqrt{(d_{A\k}-d_{B\k})^2+4|\gamma_\k|^2}\label{eq:BHC_F_Es},
\end{equation}
and their magnetic moments to be $\mu_n=g\mu_B(+\hbar S, +\hbar S)$. The two in-plane hidden components of BC ($\Omega^{a = x,y}$) for the modes are 
\begin{equation}
{\Omega}^\a_{\pm,\k}=\pm\frac{\epsilon_{\a\b}\xi}{2}\frac{\partial}{\partial k_\b}\frac{(d_{A\k}-d_{B\k})}{[(d_{A\k}-d_{B\k})^2+4 |\gamma_\k|^2]^{1/2}}\label{eq:BHC_AF_BCs},\end{equation}
which are calculated using the same pseudo-position operator as the antiferromagnet (\Sec{Sec:buckled_HC_AF}) along with the expression \eq{eq:hidden_BC_formula} in \Sec{Sec:Hidden_BC}.

Since the ordering directions for all spins are same in the BHC ferromagnet, layer symmetry must necessarily be broken either by setting $\kappa_A\neq \kappa_B$ or $D_A\neq D_B$ (\app{app:buckled_HC_F}). Breaking the layer symmetry opens a gap in the magnon bands (\subfig{Fig:buckled_HC}{f} inset), and results in a finite hidden BC distributions concentrated near the gap openings around $\k=\k_0\neq \textbf{0}$ as shown in \subfig{Fig:buckled_HC}{f}. This is in contrast to the AFM case where no gap openings were necessary and the HBCs were concentrated near $\k=\textbf{0}$. However, the symmetry of the BC distributions about the $k_y=0$ line, seen for the AFM case, still holds. 

The vertical magnon transport coefficients $\sigma^z_{x,y}$ and $D^z_{xx,xy,yy}$ arising due to the temperature gradient $(\nabla T)$ are shown in \subfig{Fig:buckled_HC}{g}, while the corresponding coefficients $\sigma^{zz}_{x,y}$ and $D^{zzz}_{xx,xy,yy}$ induced by the magnetic field gradient ($\nabla \B$) are presented in \subfig{Fig:buckled_HC}{h}. The units and the parameter values for the respective coefficients are provided in the caption of \fig{Fig:buckled_HC}.

The responses for the FM phase and AFM phase on BHC share the following similarities. The reflection symmetry $k_y \rightarrow -k_y$ makes $\sigma^z_y$, $\sigma^{zz}_y$, $D^z_{xy}$, and $D^{zzz}_{xy}$ vanish for all parameters; TRS must be broken to obtain non-zero linear coefficients; the $\nabla T$ responses saturate while $\nabla B$ coefficients grow linearly at high temperatures; and under C$_3$ symmetry, $\sigma^{z}_{x,y}$ and $\sigma^{zz}_{x,x}$ vanish while the non-linear responses satisfy $D^{z}_{xx}=D^{z}_{yy}$ (dashed line in \subfig{Fig:buckled_HC}{g}) and $D^{zzz}_{xx}=D^{zzz}_{yy}$ (dashed line in \subfig{Fig:buckled_HC}{h}).
Differences between the AFM and FM phases appear in the following key aspects -- the linear response for $\nabla \B$, $\sigma^{zz}_x$, is non-zero due to the net magnetization of the FM, even when $\kappa_A = \kappa_B$, provided layer symmetry remains broken. Unlike the AFM, the non-linear $\nabla T$ coefficients change sign at a finite temperature for the FM because the two magnon bands contribute with opposite signs. At low temperatures, the response is dominated by the lower band, while with increasing temperature, the contribution from the upper band grows. When the two contributions become equal in magnitude, the total response vanishes; beyond this point, the upper band dominates. This also explains the overall lower magnitudes of VMT responses from the FM phase compared to the AFM phase. The transport coefficients for the bilinear term in $\nabla T$, $\nabla \B$ of \eq{eq:NL_VMT} are discussed in \app{app:buckled_HC_F}.

\subsection{Vertical transport in chromium trihalides,\\ CrX$_3$ (X= I, Cl, Br)}\label{Sec:CRI3}
\begin{figure*}
\centering
\includegraphics[width=1.\textwidth]{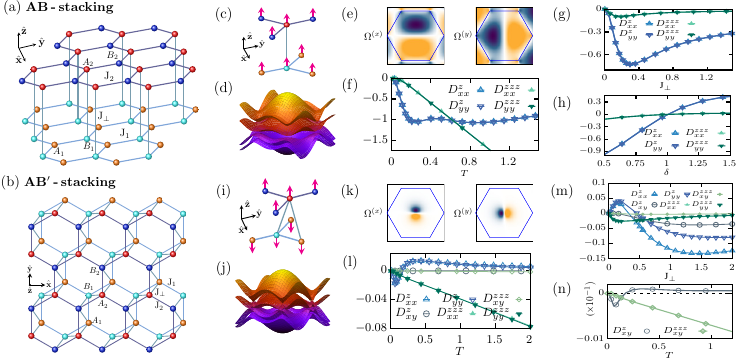}
\vspace*{-.62cm}
\caption{
\textbf{AB and AB$'$ stacked bilayer CrI$_3$}: (a) and (b) show  AB- and AB$'$-stacking geometries for bilayer CrI$_3$  with FM and AFM orders, respectively.  Bottom layer (described by parameters $\J_1,\lambda_1,\kappa_1$) and top layer (described by parameters $\J_2,\lambda_2,\kappa_2$) are vertically separated by $\xi$ and coupled with exchange couplings $\J_\perp$. Panel (c) shows magnetic ordering, (d) magnon-bands and (e) shows the hidden (in-plane) components of magnon Berry-curvature, $\Omega^{(x)}$ and $\Omega^{(y)}$, for the AB-stacked FM for a typical set of parameter values. Panels (i--k) show similar plots as (c--d) for the AB$'$-stacked AFM. The magnon-BC  within the first BZ (hexagon) are presented in units of $\xi$ with yellow (blue) regions indicating positive (negative) values. 
The linear  VMT responses, $\sigma^z_{x,y}$ and $\sigma^{zz}_{x,y}$, vanish due to time-reversal symmetry holding for both stacking geometries. 
(f) The temperature ($T$) variation of the non-linear VMT responses, $D^{z}_{xx},D^{z}_{yy},D^{zzz}_{xx}$, $D^z_{yy}$ for the AB-stacked ferromagnet with 
parameters $\J_{\perp}=0.6$, $\delta=\J_2/\J_1=\lambda_2/\lambda_1=\kappa_2/\kappa_1=0.7$.
 Responses appearing in (f) plotted by varying inter-layer coupling strength $\J_\perp$ in panel (g) for parameters $T=0.1$, $\delta=0.7$, and plotted in (h) by varying the layer-asymmetry parameter $\delta$ with  $T=0.1$, $\J_\perp=0.6$. 
Panels (l, m) show the same quantities as panels (f, g) plotted for the AB$'$-stacked AFM phase, using the parameters  $\J_{\perp}=0.04$, $\delta=0.7$ in (l), and $T=0.3$, $\delta=0.7$ in (m). A magnified view of the comparatively small but finite responses $D^z_{xy}$, $D^{zzz}_{xy}$ appearing in (l) shown in panel (n). 
A C$_3$ rotation symmetry in the AB-stacked phase causes $D^{z}_{xy}, D^{zzz}_{xy}$ to vanish and forces $D^{z}_{xx}=D^{z}_{yy}$ and $D^{zzz}_{xx}=D^{zzz}_{yy}$; no such symmetry exists for the AB$'$-stacked phase. 
The remaining parameters are set to $\J_{1}=2.2,\lambda_1=0.09,\kappa_{1}=0.01$.
}
\label{Fig:CrI3}
\end{figure*}

For our second example, we predict vertical magnon transport in van der Waals magnets, a popular class of quasi-2D materials that can be readily realized in experiments.
In particular, we demonstrate VMT in bilayer
chromium trihalides CrX$_3$ ( X = I, Cl, Br) which have become a widely adopted platform for exploring vdW physics within the experimental community. 
Both monolayer and bilayer CrX$_3$ have been successfully fabricated \cite{sivadas2018stacking} and studied, along with other few-layered materials \cite{Liu2016Van,Qi2023Fabrication,Xu2024Van}.
Several interesting phenomena including giant tunneling magneto-resistance \cite{song2018giant}, magneto-optical Kerr effect \cite{Huang2017layer,gong2017discovery}, etc., have been observed in CrX$_3$ and other layered materials. 
While most reported monolayer CrX$_3$ are ferromagnetic \cite{soriano2020magnetic}, with their magnetic moments oriented out of plane, bilayer CrI$_3$ exhibits a stacking-dependent magnetic structure \cite{chen2019direct,xu2022coexisting,zheng2018tunable,jang2019microscopic}.
The AB-stacked phase (see \subfig{Fig:CrI3}{a}) of bilayer-CrI$_3$ has been reported to host a ferromagnetic order, while the AB$'$-stacked phase shows a anti-ferromagnetic order (see \subfig{Fig:CrI3}{b}).

The general spin Hamiltonian describing the bilayer CrI$_3$ is given by
\begin{align}
    H = H_{\text{ML}}^{(1)}+H_{\text{ML}}^{(2)} + H_{\text{inter-layer}},\label{eq:bilayer_CrI3_Ham}
\end{align}
where $H_{\text{ML}}^{(l=1,2)}$ describes the Hamiltonian for the $l$-th mono-layer making up the bilayer material, and $H_{\text{inter-layer}}$ encodes the spin-spin interactions between the layers. The mono-layer spin Hamiltonian has the effective form
\begin{equation}
H_{\text{ML}}^{(l)}=-\frac{\J_l}{2}\sum_{\nn{i}{j}}\,\S^{}_{i}\cdot\S^{}_j-\frac{\lambda_l}{2}\sum_{\nn{i}{j}}\,S^{z}_{i}S^{z}_j-\kappa_l\sum_{i}\,(S^{z}_i)^2,\label{eq:CrI3_monlayer_Ham}
\end{equation}
where the first term represents the nearest-neighbor Heisenberg interactions with couplings strength $\J_l$, the second term accounts for bond anisotropy $\lambda_l$ of nearest-neighbor bonds and the last term represents single-ion anisotropies, $\kappa_l$, for the $l$-th layer. Density functional theory (DFT) estimates for these couplings are $\J_l\simeq 2.2$ meV, $\lambda_l\simeq 0.004\J_l$ and $\J_l>\lambda_l\gg \kappa_l$\cite{lado2017origin}. The magnetic moment per Cr atom in bulk CrI$_3$ is $\sim 3 \mu_B$ \cite{lado2017origin} which can be explained by a total spin value $S=3/2$. The form of the inter-layer Hamiltonian $H_{\text{inter-layer}}$  depends on the stacking arrangements of the bilayer system, and dictates whether ferro-magnetic (FM) or anti-ferromagnetic(AFM) order is favored. The ability to tune magnetic-order through stacking makes CrI$_3$ an ideal platform to study VMT in both FM and AFM phases. We study VMT for the two stacking arrangements AB and AB$'$ bellow.

\subsubsection{AB stacked bilayer CrI$_3$: Ferromagnet}\label{Sec:CrI3_FM}
In AB stacking, the A-sublattice of the top layer ($A_2$) lies directly above the B-sublattice of the bottom layer ($B_1$), while the B-sublattice of the top layer ($B_2$) is positioned above the hexagon centers of the bottom layer, as shown in \subfig{Fig:CrI3}{a}. This stacking has one interlayer nearest-neighbor bond 
and nine interlayer second-nearest-neighbor bonds. However, the coupling strength of the second closest neighbors is significantly weaker compared to that of the first closest neighbor \cite{sivadas2018stacking}. The AB-stacking shows FM ordering with all spins pointing along the $\textbf{z}$-direction(\cite{sivadas2018stacking}, see \subfig{Fig:CrI3}{a}). Therefore, the effective inter-layer Hamiltonian can be written as
\begin{align}
H_{\text{inter-layer}}=-\J_\perp\sum_{\R}\S^{(B_1)}_{\R}\cdot\S^{(A_2)}_\R,\label{eq:H_int_CrI3_F}
\end{align}
where $\R$ iterates over the Bravais-lattice vectors.
The typical value reported for the inter-layer coupling strength is $\J_\perp\simeq 0.6\,\text{meV}/\mu_B^2$\cite{sivadas2018stacking}. The layer symmetry can be broken by tuning any of the intra-layer parameters to be unequal, i.e., $\J_2 \neq \J_1$, $\lambda_2 \neq \lambda_1$, or $\kappa_1 \neq \kappa_2$.

The pseudo-$\Z$  operator for this  bilayer  model can be written as $\Z=(\xi/2)\sigma_0\otimes\text{diag}(- 1,-1,1,1)$, where $\xi$ is the vertical distance between top and bottom layers having an estimated value $\xi\simeq 3.48 $A$^\circ$\cite{Jiang2019Stacking}. Carrying out the HP transformation and LSWT analysis of the full spin Hamiltonian (\eq{eq:bilayer_CrI3_Ham}, \eq{eq:H_int_CrI3_F}), we arrive at the magnon Bloch-Hamiltonian comprising four independent modes (see \app{app:CrI3_F} for details). We  numerically compute the paraunitary matrices $\P_\k$ that diagonalize the magnon Hamiltonian, and using \eq{eq:hidden_BC_formula} evaluate the  hidden Berry curvatures ($\Omega^{(x)}$ and $\Omega^{(y)}$) for the four modes. We show the hidden BC for the energetically lowest magnon mode in \subfig{Fig:CrI3}{e}.

 Using the hidden BC and the expressions in \eq{eq:L_VMT} and \eq{eq:NL_VMT}, we compute the VMT transport coefficients for the AB-stacked FM. Since the system is time-reversal invariant in the magnon sector (see \app{app:CrI3_F}),  the linear coefficients  $\sigma^{z}_{x,y}$ and $\sigma^{zz}_{x,y}$ vanish for all values of the parameters $\J_{1,2}$, $\lambda_{1,2}$, $\kappa_{1,2}$, and $\J_\perp$.  Additional symmetries put constraints on the non-linear VMT coefficients as well. A reflection symmetry about $\textbf{y}$-axis, $k_x\rightarrow -k_x$, makes the non-linear contributions $D^{z}_{xy},D^{zzz}_{xy}$ vanish for all parameter values. Furthermore, a three-fold C$_3$ rotation symmetry constrains the remaining non-linear coefficients to follow $D^{z}_{xx}=D^{z}_{yy}$ and $D^{zzz}_{xx}=D^{zzz}_{yy}$. 
 
 We plot the temperature profile for $D^{z}_{xx},D^{z}_{yy},D^{zzz}_{xx}$ and $D^{zzz}_{yy}$ in \subfig{Fig:CrI3}{f}, which shows that coefficients, $D^{z}_{xx},D^{z}_{yy}$ arising from $\nabla T$ attain a constant value at higher temperatures, while those arising from $\nabla \B$, i.e., $D^{zzz}_{xx},D^{zzz}_{yy}$, vary linearly with $T$.  We present the dependence of $D^{z}_{xx},D^{z}_{yy},D^{zzz}_{xx},D^{zzz}_{yy}$ on the inter-layer coupling $\J_\perp$ in \subfig{Fig:CrI3}{g} for $\J_1=2.2,\lambda_1=0.09,\kappa_1=0.01$, $\J_2/\J_1=\lambda_2/\lambda_1=\kappa_2/\kappa_1=0.7$, $T=0.10$. We break layer symmetry by setting $\J_1\neq \J_2$. In the decoupled-layers limit, i.e. $\J_\perp\to0$, the said responses approach zero as expected. The responses increase with $\J_\perp$, attaining their maximum values before decreasing towards zero for very high values of $\J_\perp$.  Furthermore, we study the variation of the said responses  with layer asymmetry by defining  the parameter $\delta=\J_2/\J_1=\lambda_2/\lambda_1=\kappa_2/\kappa_1$ quantifying the degree of layer-symmetry breaking. We plot the responses  as a function of $\delta$ in \subfig{Fig:CrI3}{h} for $\J_1=2.2$, $\J_\perp=0.6$, $\lambda_1=0.09$, $\kappa_1=0.01$, and temperature $T=0.10$. We find the responses to vanish at $\delta = 1$ due to layer symmetry, and increase as $\delta$ deviates from \emph{unity}, implying that significant responses can be achieved experimentally by breaking layer symmetry.

\subsubsection{AB$'$ stacked bilayer CrI$_3$: Anti-ferromagnet}\label{Sec:CrI3_AFM}
In the AB$'$-stacked phase of bilayer CrI$_3$, an AFM order is stabilized, in which the top-layer spins point along $+\hat{\textbf{z}}$ and the bottom-layer spins along $-\hat{\textbf{z}}$, or vice versa \cite{lei2021magnetoelectric}.
Structurally, the AB$'$-stacking can be obtained from the AB-stacked geometry by a fractional lateral shift \cite{sivadas2018stacking,jang2019microscopic} of the top layer by the displacement $\R_\text{shift}=({1}/{3})\hat{\textbf{x}}$ relative to the bottom layer; see \subfig{Fig:CrI3}{b}.
Unlike in AB-stacking, a site in AB$'$-stacking connects to two nearest inter-layer neighbors (see \subfig{Fig:CrI3}{b}).
Therefore, the inter-layer Hamiltonian that stabilizes the said AFM order can be written as \cite{jang2019microscopic}
\begin{align}
H_{\text{inter-layer}}=\J_\perp\sum_{\R}&\big[\S^{(A_1)}_{\R}\cdot(\S^{(A_2)}_\R+\S^{(B_2)}_\R)\notag\\
&+\S^{(B_1)}_{\R}\cdot(\S^{(A_2)}_{\R+\r}+\S^{(B_2)}_\R)\big],\label{eq:H_int_CrI3_AF}
\end{align}
where $\R$ sums over all Bravais lattice vectors and $\r=-(1/2)\hat{\textbf{x}}-(\sqrt{3}/2)\hat{\textbf{y}}$ is a vector separating the two adjacent unit-cells connected by a bond.
The typical value of inter-layer coupling in the AB$'$-stacked phase is $\J_\perp\simeq 0.04 \text{ meV}/\mu_B^2$ \cite{sivadas2018stacking}.
The  pseudo-$\Z$ operator remains the same as in the AB-stacking case, with interlayer spacing $\xi\simeq 3.46$A$^\circ$\cite{Jiang2019Stacking}. The AFM order inherently breaks the layer symmetry, see \subfig{Fig:CrI3}{i}.

  We perform the HP and LSWT analysis of the full spin Hamiltonian (\eq{eq:bilayer_CrI3_Ham}, \eq{eq:H_int_CrI3_AF}) to obtain the magnon Bloch Hamiltonian with four modes (see \app{app:CrI3_AF}). Paraunitary matrices $\P_\k$ and the corresponding hidden Berry curvatures $\Omega^{(x,y)}$ are computed using \eq{eq:hidden_BC_formula}. The hidden BC for the lowest magnon mode is shown in \subfig{Fig:CrI3}{k}. Using the hidden BC and the expressions in \eq{eq:L_VMT} and \eq{eq:NL_VMT}, we 
compute the VMT transport coefficients for the AB$'$-stacked AFM.

Similar to the FM phase, time reversal symmetry forces the linear VMT coefficients to be zero for all parameter regimes. 
However, due to the fractional lateral shift of the top layer relative to the bottom layer, the reflection and C$_3$ rotation symmetries available in the AB-stacked FM phase are absent in the AB$'$-stacked AFM phase. Therefore, the non-linear coefficients - $D^{z}_{xy}$ and $D^{zzz}_{xy}$
which were zero in the FM phase now become finite. Also, due to absence of the C$_3$ rotation symmetry, the responses $D^{z}_{xx}\neq D^{z}_{yy}$, $D^{zzz}_{xx}\neq D^{zzz}_{yy}$ are no longer constrained to be equal. 

The temperature variation of all the non-linear coefficients are shown in \subfig{Fig:CrI3}{l}. While the large temperature behavior of the responses is similar to the AB-stacked phase, the $D^{z}_{xx}$ and $D^{z}_{yy}$ show a change of sign near $T=0.2$. Notably, the non-linear coefficients $D^{z}_{xy}$ and $D^{zzz}_{xy}$ which were earlier zero for the FM phase, are now finite but are much weaker than the $D^{z}_{xx}$ and $D^{zzz}_{yy}$ coefficients, respectively. The reason being that the magnitude of these coefficients are controlled by the scale of $\J_\perp$ which is much smaller than $\J_1$. We show a magnified version of $D^{z}_{xy}$ and $D^{zzz}_{xy}$ in \subfig{Fig:CrI3}{n}.
We present the variation of all the non-linear coefficients with $\J_\perp$ in \subfig{Fig:CrI3}{m} and find the following distinctions with the FM phase -- first, as discussed above, the breaking of C$_3$ rotation symmetry makes the $D^{z}_{xx}, D^{z}_{yy}$ responses evolve independently, second, for the value of $\J_\perp\approx 0.4 \text{ meV}/\mu_B^2$ most of the responses change sign.

Comparing the overall magnitudes of the responses between the FM and AFM phases, we find that all finite responses in the AB$'$-stacked AFM phase are an order of magnitude smaller than those in the AB-stacked FM phase. The reason is as follows. The AB$'$-stacked AFM phase can be viewed as two ferromagnetically ordered layers, with magnetizations oriented in opposite directions, coupled by an antiferromagnetic interlayer interaction (see \eq{eq:H_int_CrI3_AF}). As a consequence, the lowest two magnon bands $E_{(n=1,2),\mathbf{k}}$ (which contribute most significantly to the transport coefficients) acquire nearly identical hidden BC distributions with opposite signs, i.e., $\Omega_{1,\k} \approx -\Omega_{2,\k}$. Additionally, the magnetic moments carried by the magnons in these two bands are equal, $\mu_1 = \mu_2 = 3gS\mu_B$. Together, these features cause a partial cancellation of their contributions to VMT when summed in \eq{eq:NL_VMT}, thereby explaining the smaller response magnitudes in the AB$'$-stacked AFM phase.

Observing the above, we propose an alternate AFM phase on bilayer
CrX$_3$-type lattice geometries that achieves an order-of-magnitude larger VMT
response. This phase is constructed on the AB-stacked geometry similar to the FM phase discussed in \Sec{Sec:CrI3_FM}. However, the in-layer spin-spin interactions are tuned so that an AFM phase is stabilized on each layer as opposed to a FM phase. These two AFM layers are then anti-ferromagnetically coupled through an inter-layer interaction analogous to \eq{eq:H_int_CrI3_F}, but with $\J_\perp < 0$. The resulting lowest two bands develop hidden BC distributions that are equal in magnitude and opposite in sign, similar to the AB$'$-stacked AFM state. However, the magnetic moments $\mu_1=-\mu_2$ are of opposite signs instead of being identical. Consequently, the contributions from the two lowest bands add constructively in Eq.~\eqref{eq:NL_VMT}, yielding an order-of-magnitude enhancement of the VMT responses compared to the AB$'$-stacked AFM phase. A detailed analysis of this model and the behavior of its VMT coefficients is provided in \app{app:full_AFM}.

\section{Summary}\label{Sec:conclusion}

In this article, we have shown that atomically thin few-layered magnetic insulators can possess in-plane, or hidden, components of magnon Berry curvature (BC) in addition to the out-of-plane component typical of 2D systems. We demonstrate that these hidden components can induce magnon currents perpendicular to the plane of such quasi-2D magnets, when subjected to in-plane  temperature or magnetic-field gradients ($\nabla T$, $\nabla \B$), thereby producing vertical magnon transport (VMT). Adopting a semiclassical approach together with a dimensional crossover technique, we derive expressions for the hidden magnon BC and its contribution to vertical transport. The resulting expressions allow us to  compute VMT coefficients in terms of microscopic lattice parameters, using which we have predicted VMT in experimentally accessible van der Waals magnets, including buckled honeycomb lattice and bilayer chromium trihalides. The resulting VMT phenomena, driven by hidden magnon BC, appear to be universal, occurring in both ferro- and antiferromagnetic phases.

In closing, we offer additional insights and discuss aspects of VMT, and more importantly of hidden magnon-BC, that should prove useful when searching for these effects in experiments. For example, certain few-layered spin systems that lack an out-of-plane BC component--and thus do not exhibit conventional BC-induced transport--can still host hidden BC components and may, therefore, display VMT. The simplest example is a bilayer square lattice with multi-neighbor spin-spin interactions between the layers. Another promising platform for realizing VMT would be bilayer altermagnets similar to those discussed in \inlinecite{Cichutek2025spontaneous}. Hidden magnon-BC also opens up the possibility of 2D spintronics devices capable of sensing vertical field gradients by converting them into easily detectable in-plane currents, rather than the other way around.

Therefore, we believe that the hidden BC of magnons and the associated transport phenomena constitute a promising field to be explored both theoretically and experimentally. Doing so will not only allow us to discover and characterize layered magnetic insulators in a new light, but also holds the potential for enabling spintronic applications capable of harnessing hidden magnon BC.

\section*{Acknowledgment}
 AH acknowledges support from DST India through the ANRF grant SRG/2023/000118.

\appendix
\section{Confinement-potential approach for magnon-current}\label{app:vertical_current}
In this appendix, we provide a detailed derivation for the average magnon spin current density in \eq{eq:jz} arising due to the presence temperature ($T$) and magnetic field  ($\B$) gradients. The  magnetic field gradient ($\nabla \B$) directly appears in the semi-classical equation of motion \eq{eqm2}, determining the  anomalous velocity term and hence contributing to magnon-current. However, unlike the gradient of the magnetic field, the temperature gradient is not a body force, but rather acts as a statistical force. To obtain the magnon current induced by the temperature gradient $(\nb T)$ one can follow the confinement potential approach \cite{Ryo2011rotational}. In this approach, one considers a confining potential that vanishes inside the bulk and rises steeply at the system boundaries, thereby, preventing the magnon quasi-particles from going outside the system. With the consideration of the confining potential $U$, the semi-classical equations for the magnon wave packet \eq{eqm1}, \eq{eqm2} modify into:  
\begin{align}
\hbar\, \dot{\r}= &\;\partial_\k (E_{n\k}-\bm{\mu}_{n\k}\cdot \B)-\dot{\k}\times \bm{\Omega}\label{eq:SC1}\\
\,~\hbar \,\dot{\k}=& -\nabla(\bm{\mu}_{n\k}\cdot \B)-\nabla U= -\nabla(U+\bm{\mu}_{n\k}\cdot \B).\label{eq:SC2}
\end{align}
 Here, $E_{n\k}$ and $\bm{\Omega}_{n\k}$ denote the band dispersion and Berry curvature of the $n$-th  magnon Bloch band, respectively. The local current density corresponding to spin angular momentum carried by magnon-wave packet can be written as\cite{mook2018taking}
\begin{align*}
{\cal J}^{\rm s}_\a(\r)=\frac{1}{V} \sum_{n,\, \k}\mu^{\rm s}_{n\k}\; \vrho \; \dot{\rm r}_\a,
\end{align*}
where $\vrho\big(E_{n\k}-\bm{\mu}_{n\k}\cdot \B-U(\bm{r}),T(\r))$ is the non-equilibrium distribution function for magnons in the presence of the confining potential $U$. We are interested in the anomalous BC-dependent part of the current density ${\cal J}^{\rm s}_\a(\r)$ which, after substituting $\dot{\r}$ and $\dot{\k}$ from \eq{eq:SC1} and \eq{eq:SC2}, can be written as
\begin{align*}
{\cal J}^{\rm s}_\a(\r)=&\frac{1}{V} \sum_{n,\, \k}\mu_{n\k}^{\rm s}\,\vrho\big(E_{n\k}-\bm{\mu}_{n\k}\cdot \B(\r)-U(\bm{r}), T(\r) \big)\\
&~~~~~~~~~~\epsilon_{\a\b\c}\frac{\partial}{\partial r_\b} (U(\r)+\bm{\mu}_{n\k}\cdot \B(\r))\,{\Omega}^\c_{n\k}.\end{align*}
In the above equation, $\epsilon_{\rm abc}$ denotes the rank-3 antisymmetric Levi-Civita symbol, and repeated indices are implicitly summed. The average magnon spin current density arising due to anomalous velocity will be 
\begin{align}
 \text{j}^{\rm s}_\a=&\frac{1}{V}\int_V d^3\r~~ {\cal J}^{\rm s}_\a(\r)=\frac{1}{\hbar V} \sum_{n,\, \k}\mu_n^{\rm s} \epsilon_{\a\b\c}\;W_\b\; \Omega^\c_{n\k}\label{eq:W_omega}
\end{align} 
where we have defined
\begin{align*}
  W_\b= \frac{1}{V}\int_Vd^3\r\vrho\big(E_{n\k}-\bm{\mu}_{n\k}\cdot \B-U, T)\,\frac{\partial (U+\bm{\mu}_{n\k}\cdot \B)}{\partial r_\b}. 
\end{align*}
 First, we evaluate the integral $W_x$.
Due to the Dirac $\delta$ function nature of the confining potential gradient, i.e. $\frac{\partial {U}}{\partial x} \sim \pm \delta(x\pm \frac{L_x}{2})$, most of the contribution to the current density comes from the edges \cite{Rathor2024spin}. Therefore, we  can use the formula 
\begin{align*}
    \int dx\, F(f(x)&,g(x))\;\delta(x\pm \frac{L_x}{2})\\
    &=\int dx\, F(f(x), g(\pm\frac{L_x}{2}))\;\delta(x\pm \frac{L_x}{2})
\end{align*}
to write
\begin{align*}
W_x&=\frac{1}{V}\iint dz\,dy\Big[\int_{-\frac{L_x}{2}}^{\frac{L_x}{2}} dx \\
&~\big\{\vrho\big(E_{n\k}-\bm{\mu}_{n\k}\cdot\B(\tfrac{L_x}{2},y,z)-U, T(\tfrac{L_x}{2},y,z) \big)\\
&-\vrho\big(E_{n\k}-\bm{\mu}_{n\k}\cdot\B(-\tfrac{L_x}{2},y,z)-U, T(-\tfrac{L_x}{2},y,z) \big)\big\}\frac{\partial {U}}{\partial x}\Big]
\end{align*}
Now, in order to perform the integration over $x$, we make a change of variable $\varepsilon=U$ so that $d\varepsilon=dx\frac{\partial U}{\partial x}$  with boundary conditions   $\varepsilon=0$ at $x=0$  and $\varepsilon=\infty$ at $x=\pm L_x/2$, and  we get
\begin{align*}
W_x&=\frac{1}{V}\iint dz\,dy\Big[\int_{0}^\infty d\varepsilon \\
&\{\vrho\big(E_{n\k}-\bm{\mu}_{n\k}\cdot\B(\tfrac{L_x}{2},y,z)-\varepsilon, T(\tfrac{L_x}{2},y,z)\\
&~~~~~~~~~-\vrho\big(E_{n\k}-\bm{\mu}_{n\k}\cdot\B(-\tfrac{L_x}{2},y,z)-\varepsilon, T(-\tfrac{L_x}{2},y,z) \big)\}\Big]
\end{align*}
Using the identity $F(b)-F(a)=\int_a^b dx\,\frac{dF(x)}{dx}$, gives
\begin{align}
W_x=\frac{1}{V}\int_V d^3\r\,\int_{0}^\infty d \varepsilon\frac{\partial}{\partial x}\vrho(E_{n\k}-\bm{\mu}_{n\k}\cdot \B(\r)-\varepsilon, T(\r)).
\end{align}
The integrals $W_y$ and $W_z$ can be evaluated similarly. Substituting $W_\b$, into \eq{eq:W_omega}, we get the final expression for the average magnon spin current density, 
\begin{align}
\text{j}^{\rm s}_{\a}=\frac{1}{\hbar V^2}&\sum_{n,\, \k}\mu_{n\k}^{\rm s}\iiint_V\,d^3\r \label{eq:j_a_app}\\
&\int_{0}^\infty d \varepsilon \Big[\nb\vrho(E_{n\k}-\bm{\mu}\cdot \B(\r)-\varepsilon, T(\r))\times \bm{\Omega}_n\big]_\a\notag.
\end{align}
The $j_{a=z}$ component of the above expression is reported in \eq{eq:jz_main} of the main text. 
The magnon distribution function, $\vrho(E_{n\k}-\bm{\mu}_{n\k}\cdot \B(\r)-\varepsilon, T(\r))$, can be obtained by solving the Boltzmann transport equation (BTE)\cite{Ryo2011rotational,Kondo2022nonlinear}:
 \begin{equation}
 \frac{\partial\vrho}{\partial t} + 
 \dot{\r}\cdot\frac{\partial \vrho}{\partial \r} + 
 \dot{\k}\cdot \frac{\partial \vrho}{\partial \k}=- \frac{1}{\tau}(\vrho-\rho)  \label{eq:bte_app}. 
\end{equation}
In the above, $\tau$ is the relaxation time and $\rho\equiv \rho(E_{n\k}-\bm{\mu}_{n\k}\cdot \B(\r)-\varepsilon, T(\r))=n_B(E_{n\k}-\bm{\mu}_{n\k}\cdot \B(\r)-\varepsilon, T(\r))$  is the Bose-Einstein distribution function ($n_B(w,T) = (\exp(w/T)-1)^{-1}$) dressed with $\varepsilon$ and defined locally assuming local equilibrium holds. The expressions for $\dot{\r}$ and $\dot{\k}$ entering the above equation are taken from \eq{eqm1} and \eq{eqm2} of the main text. We look for a steady-state solution to the above equation by setting $\partial\rho/\partial t =0$.
An iterative solution \cite{kittel2018introduction} of the above equation up to $\mathcal{O}(\nabla^2)$ is given by
\begin{equation}
\vrho= \rho-\tau\,  \dot{\r}\cdot\frac{\partial \rho}{\partial \r}-{\tau}\,\dot{\k} \cdot\frac{\partial \rho}{\partial \k}+\mathcal{O}(\tau^2).\label{sol_BTE}
\end{equation}
We also consider the linear variation of temperature and magnetic field, i.e.
\begin{align*}
T(\r)= &\;T_0+r_\a(\nabla_\a T) ,~~~~~
\B(\r)=\, \B_0+r_\a(\nabla_\a \B) 
\end{align*}
We expand $\rho(E_{n\k}-\bm{\mu}_{n\k}\cdot \B(\r)-\varepsilon,T(\r))$ around the origin in terms of global distribution functions $\rho_0^\varepsilon\equiv \rho_0(E_{n\k}-\bm{\mu}_{n\k}\cdot \B_0-\varepsilon, T_0)$ as follows
\begin{widetext}
\begin{align}
\rho(E_n-\bm{\mu}_n\cdot \B(\r)-\varepsilon,T(\r))&=\,\rho_0^\varepsilon+r_\b\,\Big[\frac{\partial \rho_0^\varepsilon}{\partial T_0}(\nabla_\b T)-\frac{\partial \rho_0^\varepsilon}{\partial E_n}\mu^{\rm l}_n(\nabla_\b B_{\rm l})\Big]\label{eq:rho_ep_r}\\
&+\frac{r_\b r_\c}{2}\Big[\frac{\partial^2 \rho_0^\varepsilon}{\partial T_0^2}(\nabla_\b T)(\nabla_\c T)-2\frac{\partial^2 \rho_0^\varepsilon}{\partial T_0\partial E_n}(\nabla_\b T)\mu^{\rm r}_n(\nabla_c B_{\rm r})+\frac{\partial^2 \rho_0^\varepsilon}{\partial E_n^2} \mu^{\rm r}_n\mu^{\rm l}_n(\nabla_\b B_{\rm r})(\nabla_\c B_{\rm l})\Big]+{\cal O}(\nabla^3).\notag 
\end{align}
In the above equation, we have dropped the momentum index $\k$ from the band energy $E_{n\k}$ and the magnetic moment $\bm{\mu}_{n\k}$ for brevity. Expressing $\dot{\r}$ and $\dot{\k}$ in terms of global variables and field gradients
\begin{align}
\begin{aligned}
\hbar \,\dot{\r}_\a=&\;\frac{\partial E_{n}}{\partial k_\a}-\frac{\partial \mu^{\rm l}_{n}}{\partial k_\a}(B_{0 {\rm l}}+r_\b  \nabla_\b B_{\rm l}),~~~~~~
\hbar \,\dot{\k}_\a=-\mu^{\rm l}_{n}(\nabla_\a B_{\rm l})
\end{aligned}\label{eq:r_k}
\end{align}
and substituting the result, along with $\rho$ from \eq{eq:rho_ep_r}, into \eq{sol_BTE}, followed by a derivative of $\vrho$, we get
\begin{align}
 \nabla_\b \vrho&(E_{n}-\bm{\mu}_n\cdot \B(\r)-\varepsilon, T(\r))\notag\\
 =&\;\frac{\partial \rho_0^\varepsilon}{\partial T_0}(\nabla_\b T)-\frac{\partial \rho_0^\varepsilon}{\partial E_n}\mu^{\rm r}_n(\nabla_\b B_{\rm r})+\frac{\tau}{\hbar}\mu_n^{\rm r}(\nabla_\c B_{\rm r})\frac{\partial }{\partial k_\c}\Big[\frac{\partial \rho_0^\varepsilon}{\partial T_0}(\nabla_\b T)-\frac{\partial \rho_0^\varepsilon}{\partial E_n}\mu^{\rm l}_n(\nabla_\b B_{\rm l})\Big]\notag\\
 &-\frac{\tau}{2\hbar}\Big( \frac{\partial E_n}{\partial k_\c}-\frac{\partial \mu_n^{\rm t}}{\partial k_\c}B_{0{\rm t}}\Big)\Big[ 2\frac{\partial^2 \rho_0^\varepsilon}{\partial T_0^2}(\nabla_\b T)(\nabla_\c T)-2\frac{\partial^2 \rho_0^\varepsilon}{\partial T_0\partial E_n}\mu_n^{\rm r}\{(\nabla_\b T)(\nabla_\c B_{\rm r})+(\nabla_\c T)(\nabla_\b B_{\rm r})\}\\
 &+\frac{\partial^2 \rho_0^\varepsilon}{\partial E_n^2} \mu^{\rm r}_n\mu^{\rm l}_n \{(\nabla_\b B_{\rm r})(\nabla_\c B_{\rm l})+(\nabla_\c B_{\rm r})(\nabla_\b B_{\rm l})\}\Big]+\frac{\tau}{\hbar}\frac{\partial \mu_n^{\rm t}}{\partial k_\c}(\nabla_\b B_{\rm t})\Big[\frac{\partial \rho_0^\varepsilon}{\partial T_0}(\nabla_\c T)-\frac{\partial \rho_0^\varepsilon}{\partial E_n}\mu^{\rm r}_n(\nabla_\c B_{\rm r})\Big]\notag\\
 &+r_\c \big[\bullet\big]_{\b\c}+{\cal O}(\nabla^3).\notag  
\end{align}
We have pushed all the coefficients of $r_{\b}$ in the above equation, into $[\bullet]_{\b\c}$ as they will not contribute to the average current density.  Computing the volume integral required for \eq{eq:j_a_app}, we obtain
 \begin{align*}
\frac{1}{V}\int_V d^3\r& \nabla_\b\vrho(E_{n}-\bm{\mu}_n\cdot \B(\r)-\varepsilon, T(\r))\\
=&\;\frac{\partial \rho^\varepsilon_0}{\partial T_0}(\nabla_\b T)-\frac{\partial \rho^\varepsilon_0}{\partial E_n}\mu^{\rm r}_n(\nabla_\b B_{\rm r})+\frac{\tau}{\hbar }\mu_n^{\rm r}(\nabla_\c B_{\rm r})\frac{\partial}{\partial k_\c}\Big[\frac{\partial \rho^\varepsilon_0}{\partial T_0}(\nabla_\b T)-\frac{\partial \rho^\varepsilon_0}{\partial E_n}\mu_n^{\rm l}(\nabla_\b B_{\rm l})\Big]\\
&-\frac{\tau}{\hbar}
\Big[\frac{\partial E_n}{\partial k_\c}-\frac{\partial \mu^{\rm t}_n}{\partial k_\c}B_{0{\rm t}}\Big]\Big[\frac{\partial^2 \rho^\varepsilon_0}{\partial T_0^2}(\nabla_\b T)(\nabla_\c T)-\frac{\partial^2 \rho^\varepsilon_0}{\partial T_0\partial E_n}\mu^{\rm r}_n\{(\nabla_\b T)(\nabla_\c B_{\rm r})+(\nabla_\c T)(\nabla_\b B_{\rm r})\}\\
&+\frac{1}{2}\frac{\partial^2 \rho^\varepsilon_0}{\partial E_n^2} \mu^{\rm r}_n\mu^{\rm l}_n\{(\nabla_\b B_{\rm r})(\nabla_\c B_{\rm l})+(\nabla_\c B_{\rm r})(\nabla_\b B_{\rm l})\}\Big] +\frac{\tau}{\hbar}\frac{\partial \mu_n^{\rm r}}{\partial k_\c}(\nabla_\b B_{\rm r})\Big[\frac{\partial \rho_0^\varepsilon}{\partial T_0}(\nabla_\c T)-\frac{\partial \rho_0^\varepsilon}{\partial E_n}\mu^{\rm l}_n(\nabla_\c B_{\rm l})\Big]+
{\cal O}(\nabla^3),
\end{align*}
where we have performed the volume integral by assuming the cuboid geometry of the sample, so that 
\begin{align}
\frac{1}{V}\int_V d^3\r\;(\bullet)\equiv \frac{1}{L_xL_yL_z}\int_{-\frac{L_x}{2}}^{\frac{L_x}{2}}dx\,\int_{-\frac{L_y}{2}}^{\frac{L_y}{2}}dy\,\int_{-\frac{L_z}{2}}^{\frac{L_z}{2}}dz\,(\bullet)~~ \Longrightarrow~~ \frac{1}{V}\int_V d^3\r \;1=1,~~~~\frac{1}{V}\int_V d^3\r\; r_\a=0.
\end{align}
Next, we perform the $\varepsilon$ integrals in \eq{eq:j_a_app} using functional relations
\begin{align}
\begin{aligned}
c_0(\rho_0)\equiv~ &\frac{\partial}{\partial{E_n}}\!\int_{0}^\infty d\varepsilon \rho_0(E_n-\bm{\mu}_n\cdot\B_0-\varepsilon, T_0)=\rho_0,~~~\\
c_1(\rho_0)\equiv~ &\frac{\partial}{\partial{T_0}}\!\int_{0}^\infty d\varepsilon \rho_0(E_n-\bm{\mu}_n\cdot\B_0-\varepsilon, T_0)=(1+\rho_0)\ln(1+\rho_0)-\rho_0\ln\rho_0
\end{aligned}
\end{align}
and get  
\begin{align*}
\int_0^\infty  d\varepsilon\, \frac{1}{V}&\int_V d^3\r \nabla_\b\vrho(E_{n\k}-\bm{\mu}\cdot \B(\r)-\varepsilon, T(\r))\\
&=(\nabla_\b T)c_1-\mu_n^{\rm r}(\nabla_\b B_{\rm r})c_0+\frac{\tau}{\hbar }\mu^{\rm r}_n(\nabla_\c B_{\rm r})\frac{\partial}{\partial k_\c}\Big[(\nabla_\b T)c_1-\mu^{\rm l}_n(\nabla_\b B_{\rm l})c_0\Big]\\
&-\frac{\tau}{\hbar}
\Big[\frac{\partial E_n}{\partial k_\c}-\frac{\partial \mu^{\rm t}_n}{\partial k_\c}B_{0{\rm t}}\Big]\Big[(\nabla_\b T)(\nabla_\c T)\frac{\partial c_1}{\partial T_0}-\{(\nabla_\b T)\mu^{\rm r}_n(\nabla_\c B_r)+(\nabla_\c T)\mu^{\rm r}_n(\nabla_\b B_{\rm r})\}\frac{\partial c_1}{\partial E_n}\\
&~~~~~~+\frac{1}{2}\mu^{\rm r}_n\mu^{\rm l}_n\{(\nabla_\b B_{\rm r})(\nabla_\c B_{\rm l})+(\nabla_\c B_{\rm r})(\nabla_\b B_{\rm l})\}\frac{\partial c_0}{\partial E_n}\Big]+\frac{\tau}{\hbar}\frac{\partial \mu_n^{\rm r}}{\partial k_\c}(\nabla_\b B_{\rm r})\Big[(\nabla_\c T) c_1-\mu^{\rm l}_n(\nabla_\c B_{\rm l}) c_0\Big]+{\cal O}(\nabla^3),
\end{align*}
where $c_0(\rho_0)$ and $c_1(\rho_0)$ are functions of global density, $\rho_0\equiv n_B(E_{n\k}-\bm{\mu}_{n\k}\cdot \B_0,T_0)$. We substitute the above expression in \eq{eq:j_a_app} and write  the average magnon spin current density as
\begin{align}
\begin{aligned}
\text{j}^{\rm s}_{\a}=& \;\sigma^{\rm s}_{\a\b}(\nabla_\b T)+\sigma^{\rm s r }_{\a\b}(\nabla_\b B_{\rm r})+D^{\rm s}_{\a\b\c}\, (\nabla_\b T)(\nabla_\c T)~+D^{\rm sr}_{\a\b\c}\, (\nabla_\b T)(\nabla_\c B_{\rm r})+D^{\rm srl}_{\a\b\c}\, (\nabla_\b B_{\rm r})(\nabla_\c B_{\rm l})\label{eq:j_a_app_full}
\end{aligned}
\end{align}
where the various response tensors are given by
\begin{subequations}
\begin{align}
\sigma^{\rm s}_{\a\b}=&\;\frac{1}{\hbar V}\sum_{n,\,\k}\mu^{\rm s}_n\,\epsilon_{\a\b\e}\,\Omega_n^\e\,c_1(\rho_0),\label{eq:chi_1_app}\\
\sigma^{\rm s r }_{\a\b}=&-\frac{1}{\hbar V}\sum_{n,\,\k}\mu_n^{\rm s}\,\mu_n^{\rm r} \,\epsilon_{\a\b\e}\,\Omega_n^\e\,c_0(\rho_0),\label{eq:chi_3_app}
\end{align}
and
\begin{align}
D^{{\rm s}}_{\a\b\c}=&\;\frac{\tau}{2\hbar^2  T_0 V }\sum_{n,\,\k}\mu^{\rm s}_n\,(E_n\Omega_n^\e)\Big[\epsilon_{\a\b\e}\big(\frac{\partial c_1}{\partial k_\c}-\frac{\partial \mu_n^{\rm l}}{\partial k_\c}\frac{\partial c_1}{\partial E_n}B_{0{\rm l}}\big)+(\b \leftrightarrow \c)\Big],\label{eq:chi_2_app}\\
 D^{\rm s r l}_{\a\b\c}=&\;-\frac{\tau}{ 2\hbar^2 V}\sum_{n\k}\mu_n^{\rm s}\, \Omega_n^{\e}  \big[\epsilon_{\a\b\e}\mu_n^{\rm l}\big\{\mu_n^{\rm r}\big(2 \frac{\partial c_0}{\partial k_\c}-\frac{\partial \mu_n^{\rm t}}{\partial k_\c}\frac{\partial c_0}{\partial E_n}B_{0{\rm t}}\big)
+2 c_0 \frac{\partial \mu_n^{\rm r}}{\partial k_\c}\big\} +(\b \leftrightarrow \c) ({\rm r} \leftrightarrow \ {\rm l})\big],\label{eq:chi_4_app}\\
D^{\rm s r }_{\a\b\c}=&\;\frac{\tau}{ 2\hbar^2 V}\sum_{n,\,\k}\mu_n^{\rm s}\,\Omega_n^\e\Big[\epsilon_{\rm abe}\Big(3 \mu_n^{\rm r}\frac{\partial c_1}{\partial k_\c}-2\mu_n^{\rm r}\frac{\partial \mu_n^{\rm l}}{\partial k_\c}\frac{\partial c_1}{\partial E_n}B_{0{\rm l}}+c_1\frac{\partial \mu_n^{\rm r}}{\partial k_\c}\Big)
+(\b \leftrightarrow \c)\Big].
\end{align}
The current density reported in the \eq{eq:jz_main} in the main text is a specific case of \eq{eq:j_a_app_full} for $\a=z$ and with the simplified notation $\sigma^{\rm s(r)}_{z\a}=\sigma^{\rm s(r)}_{\a}$ and  $D^{\rm s(r)(l)}_{z\a\b}=D^{\rm s(r)(l)}_{\a\b}$.
\end{subequations}
\end{widetext}
\section{Hidden Berry curvatures of bosonic BdG system}\label{app:BdG_BC}
Magnon Bloch-Hamiltonian generically have a BdG form \cite{Matsumoto2014Thermal,Kondo2020Non} and independent magnon modes are obtained using paraunitary transformations $\bm{\para}_\k$. Consequently, the Berry connections for these BdG-systems are given by \cite{Matsumoto2014Thermal}
\begin{equation}
    {\cal A}_n^\a= -i\,\Big[\Sigma_z\bm{\para}^\dagger_{\k}\,\Sigma_z\frac{\partial \bm{\para}_{\k}}{\partial {k_\a}}\Big]_{nn}. 
\end{equation}
Subsequently, the Berry curvature is defined as
\begin{align*}
  {\Omega}_{n\k}^\a\equiv (\nabla_\k\times {\cal A}_n)^\a= &-i\,\epsilon_{\a\b\c}\frac{\partial}{\partial {k_\b}}\Big[\Sigma_z\bm{\para}^\dagger_{\k}\,\Sigma_z\frac{\partial \bm{\para}_{\k}}{\partial {k_\c}}\Big]_{nn}
\end{align*}
The form of the in-plane components of the magnon BC, $\Omega^{(x)}$ and $\Omega^{(y)}$ will therefore be
\begin{align}
  {\Omega}_{n\k}^\a=-i\,\epsilon_{\a\b}\Big[\Big(\Sigma_z\frac{\partial \bm{\para}^\dagger_{\k}}{\partial {k_\b}}\,\Sigma_z\frac{\partial \bm{\para}_{\k}}{\partial {k_z}}\Big)
 -\Big(\Sigma_z\frac{\partial \bm{\para}^\dagger_{\k}}{\partial {k_z}}\,\Sigma_z\frac{\partial \bm{\para}_{\k}}{\partial {k_\b}}\Big)\Big]_{nn},\label{eq:HBC}
  \end{align}
where $\epsilon_{\a\b}=\epsilon_{\a\b z}$ is Levi-Civita  tensor of rank 2. We apply the above expression to compute the in-plane components of Berry curvatures for quasi-2D system similar to that given in \eq{eq:Ham_quasi2D}. As discusses in the main text, for  $L$-layered systems for which 
\begin{equation}
\bm{\para}=\para_{k_z}\para_{\k}~~\text{with}~~\para_{k_z}=I_2\otimes \text{diag}(e^{ik_z \xi_1}{\cal I}_1,\cdots,e^{ik_z \xi_L}{\cal I}_L),\label{eq:para_product_app}
\end{equation}
where ${\cal I}_1,\cdots,{\cal I}_L$ are the identity matrices for each layer. Substituting the said product decomposition for $\P$ into \eq{eq:HBC}, we obtain
\begin{align}
\Omega_{n\k}^\a=-i\epsilon_{\a\b}\Big[ \Sigma_z&\frac{ \partial \para^\dagger_\k}{\partial k_\b}\big(\para^\dagger_{k_z}\Sigma_z\frac{ \partial\para_{k_z}}{\partial k_z}\big) \para_\k\notag\\
&~~-\Sigma_z\para^\dagger_\k\big(\frac{ \partial \para_{k_z}^\dagger}{\partial k_z}\Sigma_z \para_{k_z}\big)\frac{\partial\para}{\partial k_\b}\Big]_{nn}\label{eq:app_B_eq}
\end{align} 
Using \eq{eq:para_product_app}, we can simplify the following terms appearing in the above expression: 
\begin{equation}
\big(\para^\dagger_{k_z}\Sigma_z\frac{ \partial\para_{k_z}}{\partial k_z}\big) =i \Z ~~~\text{and}~~~\frac{ \partial \para_{k_z}^\dagger}{\partial k_z}\Sigma_z \para_{k_z}=-i\Z,\notag
\end{equation}
where $\Z=\Sigma_z\otimes \text{diag}( \xi_1 {\cal I}_1,\cdots,\xi_L{\cal I}_L)$. Utilizing these simplifications, we re-express \eq{eq:app_B_eq} to get 
\begin{align*}
\Omega^\a_{n\k}=&\epsilon_{\a\b}\Big[ \Sigma_z\frac{ \partial \para^\dagger_\k}{\partial k_\b} \Z\,\para_\k\Big]_{nn}+\epsilon_{\a\b}\Big[ \Sigma_z\para^\dagger_\k\,\Z\frac{\partial\para}{\partial k_\b}\Big]_{nn},
\end{align*}
which simplifies to an elegant expression for the hidden magnon Berry curvatures (HMBC)
\begin{align}
\Omega^\a_{n\k}=\epsilon_{\a\b}\frac{ \partial}{\partial k_\b}\Big[ \Sigma_z \para^\dagger_\k\, \Z\,\para_\k\Big]_{nn},\label{eq:HBC_formula_app}
\end{align}
which we report in \eq{eq:hidden_BC_formula} of the main text.

We show the following properties to hold for HMBC below:\\

\noindent 1. {\it Hidden Berry curvatures do not depend upon the absolute vertical positions of the layers but on their relative vertical positions only}. This can be proved by considering a position operator created by shifting the origin along the $\text{z}$-axis by $\xi_0$, i.e., $\Z'=\Z-\xi_0 \Sigma_z$. Due to the paraunitary property $\para^\dagger_\k\Sigma_z \,\para_\k=\Sigma_z$,  we will have $\Sigma_z \para^\dagger_\k\Z'\,\para_\k=\Sigma_z \para^\dagger_\k\Z\,\para_\k-\xi_0 \Sigma_0$ with $\Sigma_0$ being the identity matrix. The constant matrix $\xi_0 \Sigma$ vanishes after taking the $k$-derivative in \eq{eq:HBC_formula_app}, thus keeping the hidden BC unchanged.\\

\noindent 
2. {\it Hidden Berry curvatures for a perfectly flat system are zero}. This follows from the previous point. Since the relative vertical positions of the sites in a flat layer are zero, so are the hidden BCs.\\

\noindent 3. {\it Hidden Berry-curvatures for a system of decoupled-layers are zero}. In a trivial case, when all the layers are decoupled, hidden-BCs will vanish identically over the entire BZ. This can be understood as follows. For such systems, the paraunitary matrix with a change of basis can be written as $\para_\k=\text{diag}(\para_{1\k},\para_{2\k},\cdots ,\para_{L\k})$ where $\para_{l\k}$ are the paraunitary  eigen-vector matrices of the $l$-\text{th} layer satisfying $\para^\dagger_{l\k}\Sigma^{(l)}_z \para_{l\k}=\Sigma^{(l)}_z$, where $\Sigma^{(l)}_z= {\cal I}_l \otimes\sigma_z$. The pseudo-$\Z$ operator in this new basis will be $\Z=\text{diag}(\xi_1 {\cal I}_1,\cdots\cdots, \xi_L {\cal I}_L)\otimes\,\sigma_z$ . The above expressions together imply
\[\para^\dagger_\k\Z\para_\k = \text{diag}(\xi_1 {\cal I}_1,\cdots\cdots, \xi_L {\cal I}_L)\otimes\,\sigma_z,\] meaning
$\para^\dagger_\k\Z\para_\k $ is $\k$-independent, thereby leading to zero hidden-BCs (see \eq{eq:HBC_formula_app}).\\

\noindent 4. {\it Breaking the layer symmetry is a necessary condition to obtain a nonzero hidden Berry curvature}. To support this argument, we show that the hidden BCs vanish when all the layers are identical. In such a scenario, the system is symmetric under inter-change of any two layers. This implies that the operator exchanging the layer $l$ with $l'$ and represented by the permutation matrix $\Sllp$, will commute with the magnon Hamiltonian $\H_\k$, as well as with $\Sigma_z \H_\k$, i.e. $[\Sllp,~\H_\k]=[\Sllp,~~ \Sigma_z\H_\k]=0$. The said commutations will also hold for all choices of $l, l'$. The paraunitary matrix $\para_\k$ which diagonalizes $\Sigma_z \H_\k$,  can  be chosen to be a eigenvector matrix of $S_{ll'}$ (for all $l,l'$), i.e.
\begin{equation}
   S_{ll'}\para_\k = \para_\k \Lllp,~~~~~\forall~~ l,l'\label{eq:app_B_Lambda}
\end{equation}
where $\Lllp$ is a diagonal matrix with entries $\pm 1$ since permutation matrices have only $\pm 1$ eigen values.  Using \eq{eq:app_B_Lambda}, we have
\begin{equation}
    \para^\dagger_\k \Z \para_\k =    \Lllp\para^\dagger_\k \big(S_{ll'}^\dagger\,\Z\, S_{ll'}\big)\para_\k  \Lllp,~~~~~\forall ~~l,l'\label{eq:PdZP=LPdSdZSPL}
\end{equation}
where $\Z=\sigma_z\otimes\,\text{diag}(\xi_1,\xi_2,\cdots, \xi_L)\otimes {\cal I}$
is the pseudo-$\Z$ operator and  ${\cal I}$ is the identity matrix for each layer. 
Since according to \eq{eq:HBC_formula_app} we only need the diagonal elements of $\para^\dagger_\k \Z\para_\k$, we can simplify \eq{eq:PdZP=LPdSdZSPL} as 
\begin{align}
    \big[\para^\dagger_\k \Z \para_\k\big]_{nn} =  &\;  [\Lllp]_{nn}[\Lllp]_{nn}\big[\para^\dagger_\k \,S_{ll'}^\dagger\,\Z\, S_{ll'}\,\para_\k\big]_{nn}\nonumber\\
   =&\; \big[\para^\dagger_\k \, S_{ll'}^\dagger\,\Z\, S_{ll'}\,\para_\k\big]_{nn},~~~~~\forall ~~l,l'\label{eq:PdZP=PdSdZSP},
\end{align}
where we have eliminated $\Lllp$.
The action of the permutation matrix $\Sllp$ on the pseudo-$\Z$ operators results in a permuted operator  $\Z^{(l\leftrightarrow l')}\equiv S_{ll'}^\dagger\,\Z\, S_{ll'}$ having the explicit form
\begin{equation}
\Zllp=
\sigma_z \otimes 
\text{diag}(\xi_1,\cdots,\!\!\!\!\underbrace{\xi_{l'}}_{l\text{-th element}},\cdots,
\!\!\!\!\underbrace{\xi_{l}}_{l'\text{-th element}},\dots,\xi_L)\otimes \mathcal I.    
\end{equation}
\Eq{eq:PdZP=PdSdZSP} when written in terms of $\Zllp$, takes the form
\begin{align}
    \big[\para^\dagger_\k \Z\para_\k\big]_{nn}=    \big[\para^\dagger_\k \Zllp\para_\k\big]_{nn}.
\end{align}
Summing up the above equation  $L$ times for the permutations, $S_{12},S_{2\,3},\cdots,S_{L-1\,L},S_{L\,1}$, we get
\begin{align}
 \big[\para^\dagger_\k \Z \para_\k\big]_{nn} =&\;\frac{1}{L}\big[\para^\dagger_\k \big(\Z^{(1\leftrightarrow 2)}+\Z^{(2\leftrightarrow 3)}\notag\\
 &~~~~~~~+\cdots +\Z^{(L-1\leftrightarrow L)}+\Z^{(L\leftrightarrow 1)} \big)\para_\k \big]_{nn}\notag \\
=&\frac{1}{L}\big[\para^\dagger_\k \big(\sigma_z\otimes (\xi_1+\xi_2+\cdots+\xi_L){\cal I}_L\otimes {\cal I} \big)\para_\k \big]_{nn}\notag\\
=&\frac{1}{L}(\xi_1+\xi_2+\cdots+\xi_L)\big[\para^\dagger_\k \big(\sigma_z\otimes {\cal I}_L\otimes {\cal I} \big)\para_\k \big]_{nn}\notag\\ 
 =&\frac{1}{L}(\xi_1+\xi_2+\cdots+\xi_L)\big[\para^\dagger_\k \,\Sigma_z\,\para_\k\big]_{nn}\notag\\ 
 =&\frac{1}{L}(\xi_1+\xi_2+\cdots+\xi_L)\big[\Sigma_z\big]_{nn}\,\notag \end{align}
yielding a constant at the end, which when substituted in \eq{eq:HBC_formula_app} leads to a {\it zero} hidden Berry curvature for all momenta $\k$ and all bands $n$.
\section{Model specific details}\label{app:models}
\subsection{Buckled-honeycomb anti-ferromagnet}\label{app:buckled_HC_AF}
For the spin Hamiltonian (\eq{eq:BHC_spin_Ham} of the main text) describing the buckled honeycomb lattice, an AFM order is obtained when $\J_{1,2}>0$. The easy-axis anisotropies along the $z$-axis, $\kappa_{A,B}$ enforce spin-ordering parallel to the $\hat{\textbf{z}}$-direction. The Holstein-Primakoff (HP) transformations for this anti-ferromagnetic ordering are provided in the main text near \eq{eq:Fourier_transform}. Using  these HP transformations, we represent Heisenberg interactions between spins in  terms of boson operators up to quadratic order as follows;
\begin{align*}
\S^{(A)}_i\cdot \S^{(B)}_j =-S^2+ S(a^\dagger_i a_i+b^\dagger_j b_j)+S(a_i b_j+a^\dagger_i b^\dagger_j),
\end{align*}
the DM-interaction terms as
\begin{align*}
(\S^{(A)}_i\times \S^{(A)}_j)^z =&-iS(a^\dagger_i a_j-a_i a^\dagger_j),\\
(\S^{(B)}_i\times \S^{(B)}_j)^z =&\,iS(b^\dagger_i b_j-b_i b^\dagger_j),
\end{align*}
and similarly, the easy-axis an-isotropy terms as
\begin{align*}
(S^{(A)z}_i)^2 =S^2-2S\,a^\dagger_i a_i , ~~~~(S^{(B)z}_j)^2 = S^2- 2S\,b^\dagger_j b_j. \end{align*}
We now substitute the above expressions  in the spin Hamiltonian \eq{eq:BHC_spin_Ham} to get
\begin{align}
\begin{aligned}
H\simeq ~H_{0}&+S(2\J_1+\J_2+2\kappa_A)\;\sum_{i} a^\dagger_{i}\,a_{i}\\
&+S(2\J_1+\J_2+2\kappa_B)\;\sum_{i} b^\dagger_{i}\,b_{i}\\
&+S \J_1 \sum_{\langle i,j\rangle\in \text{ slant}} \!\big( a_i b_j+a^\dagger_i b^\dagger_j\big)\\
&+S \J_2 \sum_{\langle i,j\rangle\in \text{vertical}}\! \big( a_{i}b_{j}+a^\dagger_{i}b^\dagger_{j}\big)\\
&-iSD_A \sum_{\nnn{i}{j}} \!\big( a^\dagger_i a_j-a_i a^\dagger_j\big)\\
&+iSD_B \sum_{\nnn{i}{j}} \!\big( b^\dagger_i b_j-b_i b^\dagger_j\big).
\end{aligned}
\end{align}
The Hamiltonian $H_{0}$ on the right hand side of the above equation is ${\cal O}(S^2)$ and gives the classical energy of the magnetically ordered ground state, and the remaining parts constitute the magnon Hamiltonian $H_\text{magnon}$ describing the quantum magnon excitations. Expressing the magnon Hamiltonian in terms of Fourier space operators (defined in \eq{eq:Fourier_transform}), we obtain
\begin{align*}
\begin{aligned}
H_\text{magnon}=\sum_{\k } d_{A\k} \;a^\dagger_{\k}\,a_{\k}+d_{B\k}\;b^\dagger_{\k}\,b_{\k}+\gamma_{\k}^* a_{\k}b_{\tminus\k}+\gamma_{\k}a^\dagger_{\k}b^\dagger_{\tminus\k}\big.
\end{aligned}
\end{align*}
where the symbols $d_{A\k},d_{B\k}$ and $\gamma_\k$ are defined in the main text (below \eq{eq:Ham_BHC_AF}).
The matrix form of the magnon Hamiltonian is given in \eq{eq:Ham_BHC_AF}. 

Time-reversal symmetry is present in the
magnon sector for this model when $D_A=D_B$ and is broken otherwise.
This can be explicitly checked by constructing the time reversal operator ${\cal T}=\sigma_x\otimes \,\sigma_0 K$, and showing
${\cal T} \H_\k{\cal T}^{-1}= \H_{-\k}$ when $D_A=D_B$.
In the definition for ${\cal T}$, $\sigma_0$ represents the $2 \times 2$ identity matrix, $\sigma_{x}$ is a Pauli matrix and $K$ denotes complex conjugation operator. 
The model also possesses a reflection symmetry $k_y\rightarrow -k_y$. It furthers, exhibits a three fold rotation symmetry C$_3$ about the $z$-axis for $\J_1=\J_2$. Independent magnon modes can be obtained by diagonalizing $\H_\k$ using the paraunitary matrix
\begin{equation}
\para_\k=\!\!\left[\!\!\begin{array}{cccc}
\cosh \frac{\beta_\k}{2}& 0&0 &\sinh \frac{\beta_\k}{2}e^{i\alpha_\k}\\
0&\cosh\frac{\beta_{\textbf{-}\k}}{2} &\!\!\!\sinh \frac{\beta_{\textbf{-}\k}}{2}e^{i\alpha_{\textbf{-}\k}}&0\\
0&\!\!\sinh\frac{\beta_{\textbf{-}\k}}{2}e^{\textbf{-}i\alpha_{\textbf{-}\k}}&\cosh \frac{\beta_{\textbf{-}\k}}{2} &0\\
\sinh \frac{\beta_{\k}}{2}e^{\textbf{-}i\alpha_{\k}} &0&0&\cosh \frac{\beta_{\k}}{2}
\end{array}\!\!\right]\notag 
\end{equation} 
\begin{figure}
    \centering
\includegraphics[width=.48\textwidth]{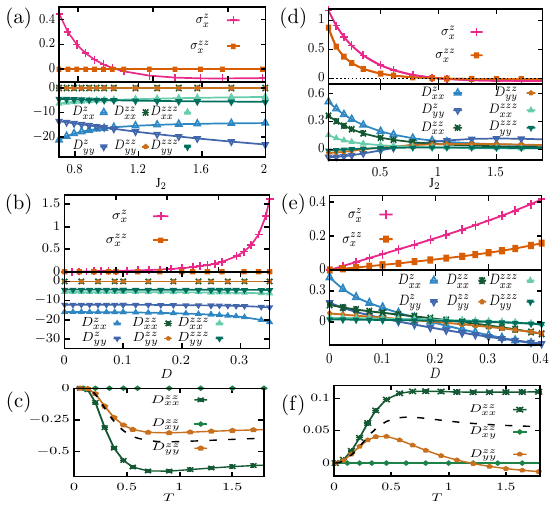}
    \caption{Panel (a) shows the VMT responses as functions of $\J_2$ (vertical-bond coupling) for the AFM state, using parameters $\J_1=1.0$, $D_A=-D_B=0.35$, $\kappa_A=\kappa_B=0.01$, and $T=0.4$. Panel (b) presents the corresponding responses as functions of $D$ (with $D=D_A=-D_B$) for $\J_1=1.0$, $\J_2=0.7$, $\kappa_A=\kappa_B=0.01$, and $T=0.4$. Panel (c) displays $D^{zz}_{xx}$, $D^{zz}_{xy}$, and $D^{zz}_{yy}$ for the AFM state as functions of temperature, using $\J_1=1.0$, $\J_2=0.7$, $D_A=-D_B=0.35$, and $\kappa_A-\kappa_B=0.2$. Panel (d) shows the VMT responses for the FM state as functions of $\J_2$, with parameters $\J_1=1.0$, $D_A=D_B=0.1$, $\kappa_A=0.1$, $\kappa_B=0.01$, and $T=0.4$. Panel (e) plots the same responses versus $D$ (where $D=D_A=D_B$) for $\J_1=1.0$, $\J_2=0.7$, $\kappa_A=0.1$, $\kappa_B=0.01$, and $T=0.4$. Panel (f) shows $D^{zz}_{xx}$, $D^{zz}_{xy}$, and $D^{zz}_{yy}$ for the FM state as functions of temperature for $\J_1=1.0$, $\J_2=0.7$, $D_A=D_B=0.1$, $\kappa_A=0.1$, and $\kappa_B=0.01$.}
    \label{Fig:BHC_app}
\end{figure}
where $\alpha_\k=\text{arg}(\gamma_\k)$ and $\tanh (\beta_\k)=-\frac{2|\gamma_\k|}{(d_{A\k}+d_{B\textbf{-}\k})}$. It can be easily checked that $\para_\k$ satisfy the the para-uniatry conditions $\para^\dagger_\k \Sigma_z \para_\k=\Sigma_z$. The resulting energies of the four magnon modes are $E_{-,\k}, E_{+,\k},-E_{-,\k},-E_{+,\k}$, where $E_{\pm,\k}$ are reported in \eq{eq:band_dispersion_BHC_AF}.
We compute the hidden or in-plane components of magnon-Berry curvatures using pseudo-position operator $\Z= \Sigma_z\otimes \text{diag}(\frac{\xi}{2},-\frac{\xi}{2})$ and \eq{eq:hidden_BC_formula}. The resulting Berry curvature for the four magnon bands are $\Omega^\a_\k,\Omega^\a_{-\k},-\Omega^\a_\k,-\Omega^\a_{-\k}$, where
\begin{align}
    \Omega_\k^\a=&\frac{\xi \epsilon_{\a\b}}{2}\frac{\partial }{\partial k_\b}\big(\cosh \beta_\k\big),~~~~\a,\b= x,y\\\
    \Omega^{(z)}_\k=&\frac{1}{2}\sinh \beta_\k\frac{\partial (\beta_\k, ~\alpha_\k)}{\partial (k_x,~k_y)}.
\end{align}
The out-of-plane component, $\Omega^{(z)}_\k$ is computed using \eq{eq:BCmunu}. It is worth noting that unlike the out-of-plane component $\Omega^{(z)}_\k$, the in-plane components do not depend upon $\alpha_\k$. By substituting $\beta_\k$ in terms of $d_{A\k}$, $d_{B\tminus\k}$ and $\gamma_\k$ we get the expression \eq{eq:BC_BHC_AF} for the hidden magnon BCs used in the main text.

The vertical magnon transport (VMT) responses for the buckled honeycomb AFM were plotted as a function of temperature $T$ in \subfigs{Fig:buckled_HC}{d}{e} of the main text. Here, we plot them by varying $\J_2$ and also by varying $D(=D_{A}=-D_B)$ in \subfigs{Fig:BHC_app}{a}{b}, respectively. We also plot the bilinear responses--$D^{zz}_{xx}$, $D^{zz}_{xy}$, $D^{zz}_{yy}$ (defined in \eq{eq:chi_5})--arising from the simultaneous application of temperature gradient $(\nabla T)$ and magnetic field gradient $(\nabla \B)$  in \subfig{Fig:BHC_app}{c}. 
\subsection{Buckled-honeycomb ferromagnet}\label{app:buckled_HC_F}
The magnon Hamiltonian for the ferromagnetic order on the honeycomb lattice, can also be derived using steps similar to those used for the antiferromagnetic case discussed in \app{app:buckled_HC_AF}. 
The form of the magnon Bloch Hamiltonian  $h_\k$ is given in \eq{eq:hk_F}. For the FM case, time-reversal symmetry is preserved when $D_A=-D_B$ and broken otherwise.
The time reversal operator for this case is ${\cal T}=\sigma_0 K$ where $\sigma_0$  is $2 \times 2$ identity matrix and $K$ is the complex conjugation operator. Thus, for $D_A=-D_B$ we have ${\cal T} h_\k{\cal T}^{-1}= h_{-\k}$. The reflection symmetry $k_y\rightarrow -k_y$ as well as the three-fold rotational symmetry C$_3$ (present when $\J_1=\J_2$), also holds in this case. The two independent magnon modes can be obtained by diagonalizing $h_\k$ (defined in \eq{eq:hk_F} ) using an unitary matrix $U_\k$;
\begin{equation}
    U_\k=\bigg[\begin{array}{cc}
       \cos \frac{\beta_\k}{2}& \rminus\sin \frac{\beta_\k}{2}e^{i\alpha_\k}\\
       \sin \frac{\beta_\k}{2}e^{-i\alpha_\k}&\cos \frac{\beta_\k}{2}
    \end{array}\bigg]\notag
\end{equation}
where $\tan \beta_\k= \frac{2|\gamma_\k|}{(d_{A\k}-d_{B\k})}$ and $\alpha_\k=\arg(\gamma_\k)$.
The position operator is $\Z=\text{diag}(\frac{\xi}{2},-\frac{\xi}{2})$. The dispersion relations for the two magnon bands are given in \eq{eq:BHC_F_Es}. The hidden Berry curvatures for the two magnon bands are $\Omega^{\a}_\k,-\Omega^{\a}_\k$, where 
\begin{equation}
  \Omega^\a_\k=\frac{\xi \epsilon_{\a\b}}{2}\frac{\partial }{\partial k_\b}\big(\cos \beta_\k \big),~~~~~ \a=x,y. 
\end{equation}
When $d_{A\k}=d_{B\k}$, the two layers become identical and the hidden BCs vanish over entire Brillouin zone (BZ). The layer symmetry can be broken either by setting $\kappa_A\neq \kappa_B$ or by setting $D_A\neq D_B$.
The temperature dependence of the vertical magnon transport (VMT) responses for the FM case are reported in \subfigs{Fig:buckled_HC}{g}{h} of the main text. We plot their variation with $\J_2$ and with $D(=D_A=D_B)$ in \subfig{Fig:BHC_app}{d} and in \subfig{Fig:BHC_app}{e}, respectively. We also plot the VMT responses--$D^{zz}_{xx},D^{zz}_{xy}$, $D^{zz}_{yy}$ (defined in \eq{eq:chi_5})--arising from the simultaneous application of temperature gradient $(\nabla T)$ and magnetic field gradient $(\nabla \B)$ in \subfig{Fig:BHC_app}{f}.
\subsection{AB-stacked bilayer CrI$_3$}\label{app:CrI3_F}
The spin Hamiltonian for AB-stacked bilayer CrI$_3$ is given by \eq{eq:bilayer_CrI3_Ham} to \eq{eq:H_int_CrI3_F} of the main text. The easy axis anisotropy terms in  \eq{eq:CrI3_monlayer_Ham} favor all the spins in each layer to align along the $z$-axis, and the inter-layer Hamiltonian in \eq{eq:H_int_CrI3_F} stabilizes a parallel alignment between the two layers, resulting in a ferromagnetic ordering (see \subfig{Fig:CrI3}{c}). The HP transformations for the spins on sub-lattices $A_{1,2}$ and $B_{1,2}$ on both the layers are
\begin{align}
\begin{split}
  \S^{\scriptscriptstyle (A_l)-}_i=& \;(2S-a^\dagger_{li} a_{li})^{\frac{1}{2}}a^\dagger_{li},\\
   \S^{\scriptscriptstyle (A_l)+}_i=&\;a_{li}(2S-a^\dagger_i a_{li})^{\frac{1}{2}},\\
   \S^{\scriptscriptstyle (A_l)z}=&\;S-a^\dagger_{li} a_{li},
\end{split}
\Bigg|
\begin{split}   
   \S^{\scriptscriptstyle (B_l)-}_i=& \;(2S-b^\dagger_{li} b_{li})^{\frac{1}{2}}b^\dagger_{li},\\
   \S^{\scriptscriptstyle (B_l)+}_i=&\;b_{li}(2S-b^\dagger_{li} b_{li})^{\frac{1}{2}},\\
   \S^{\scriptscriptstyle (B_l)z}=&\;S-b^\dagger_{li} b_{li}
\end{split} \notag\label{eq:BHC_FM_HP}    
\end{align}
for $l=1,2$. Following similar steps as those performed for the buckled honeycomb AFM (\app{app:buckled_HC_AF}), we obtain the magnon Hamiltonian $H_\text{magnon}$ for this model in terms of the real space operators $a_{1,2i},b_{1,2i}$ and consequently use their Fourier transforms
\begin{equation}
\bigg[\begin{array}{cc}
a_{l\,\k}  \\b_{l\,\k}  \end{array}\bigg]
=\sum_{\R_i}e^{-i\k\cdot \R_i} \bigg[\begin{array}{cc}e^{-i\k\cdot \r_A} a_{l\,i}  \\[1pt]e^{-i\k\cdot \r_B} b_{l\,i} \end{array}\bigg],~~~l=1,2\label{eq:Fourier_transform_app}
\end{equation}
to arrive at 
\begin{equation}
H_\text{magnon}=\sum\limits_{\k} \bm{\phi}^\dagger_\k h_\k\bm{\phi}_{\k},
~~ \bm{\phi}_\k\equiv [a_{1\,\k}~~b_{1\,\k}~~a_{2\,\k}~~b_{2\,\k}]^T.  \label{eq:app_BCS_magH}  
\end{equation}
Due to the overall FM order, the Bloch Hamiltonian in the above equation has been reduced to a non-BdG form since $\Delta_\k=\bm{0}$, 
where
\begin{align}
h_\k=
\begin{bmatrix}
~~d_{1}&\rminus \frac{\J_1}{2}\gamma^{}_{\k}&0&0\\
\rminus \frac{\J_1}{2}\gamma^{*}_{\k}&d_{1}+\J_\perp&\rminus\J_\perp&0\\
0&\rminus\J_\perp&d_{2}+\J_\perp&\rminus\frac{\J_2}{2} \gamma_{\k}\\
0&0&\rminus\frac{\J_2}{2} \gamma^{*}_{\k}&d_{2}
\end{bmatrix}\label{eq:app_ABS_coe_mag_H}
\end{align}
with $d_{1,2}=(\frac{3}{2}\J_{1,2}+\frac{3}{2}\lambda_{1,2}+2\kappa_{1,2})$ and $\gamma_\k=\exp (i{k_y}/{\sqrt{3}})+2\cos(k_x/2)\exp (-i{k_y}/{2\sqrt{3}})$. Here, $\J_{1,2},\lambda_{1,2},\kappa_{1,2}$ are monolayer  parameters of the bottom and top layers, respectively, and $\J_\perp$ is the coupling strength of the inter-layer bonds. Since the spin Hamiltonian of this model includes only Heisenberg exchange and single-ion anisotropy, and no Dzyaloshinskii-Moriya interactions, the resulting magnon Hamiltonian preserves time-reversal symmetry. The time-reversal operator for this model is given by ${\cal T} = \sigma_0 \otimes \sigma_0 K$, where $\sigma_0$ is the $2\times2$ identity matrix and $K$ denotes complex conjugation operator. Consequently, the magnon Hamiltonian satisfies ${\cal T} h_\k {\cal T}^{-1}=h_{\tminus\k}$.
\begin{figure}
    \centering
\includegraphics[width=0.99\linewidth]{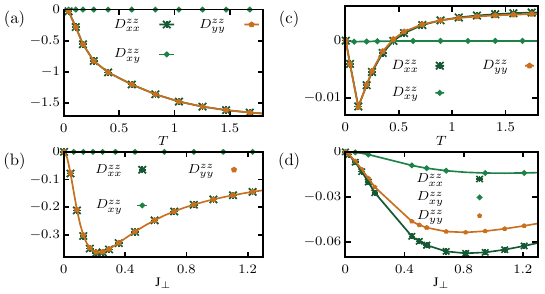}
    \caption{The bilinear VMT responses generated by simultaneous application of temperature gradient $(\nabla T)$  and magnetic field gradient-- namely $(\nabla \B)$--$D^{zz}_{xx},D^{zz}_{xy}$ and $D^{zz}_{yy}$--are shown for the AB-stacked phase of CrI$_3$ as functions of temperature in panel (a) using parameters $\J_{\perp}=0.6$, $\delta=0.7$, and as functions of interlayer coupling $\J_\perp$ in panel (b), using $T=0.1$, $\delta=0.7$. Corresponding responses for the AB$'$-stacked phase are as functions of  temperature in panel (c), using $\J_{\perp}=0.04$, $\delta=0.7$ and as functions of $\J_\perp$ in panel (d) using $T=0.3$, $\delta=0.7$. All ther parameters are fixed at $\J_1=2.2,\lambda_1=0.09 $ and $\kappa_1=0.01$.}
    \label{Fig:CrI3_bilinear_VMT}
\end{figure}

The vertical magnon transport responses arising due to pure temperature gradient $(\nabla T)$ and magnetic field gradient ($\nabla \B$) are
reported in \fig{Fig:CrI3} of the main text. 
Here, we plot the temperature-dependence of the VMT responses--$D^{zz}_{xx},D^{zz}_{xy}$, $D^{zz}_{yy}$--induced when both temperature gradient and magnetic field gradient are applied simultaneously in \subfig{Fig:CrI3_bilinear_VMT}{a}. The same responses are also presented as functions of $\J_\perp$ in \subfig{Fig:CrI3_bilinear_VMT}{b}.
\subsection{AB$'$-stacked bilayer CrI$_3$}\label{app:CrI3_AF}
The AB$'$-stacking geometry corresponds to a fractional lateral shift of top layer by $\R_\text{shift}=({1}/{3})\hat{\textbf{x}}$ from AB-stacking geometry, see \fig{Fig:CrI3}. In the AB-stacking, site $A_2$ lies exactly at the top of site $B_1$. 
Therefore, in AB$'$-stacking we have, $\r_{A_2B_1}=\frac{a}{3}\hat{\textbf{x}}+\xi\, \hat{\textbf{z}} $.
The new relative positions of the sites within the unit cell due to the lateral shift $\R_\text{shift}$ are given by
\begin{align*}
\r_{B_2A_1}=&\,-\frac{a}{6}\hat{\textbf{x}}-\frac{a}{2\sqrt{3}}\hat{\textbf{y}}+\xi\, \hat{\textbf{z}} \\  \r_{B_2A_1}=&\,-\frac{a}{6}\hat{\textbf{x}}+\frac{a}{2\sqrt{3}}\hat{\textbf{y}}+\xi\, \hat{\textbf{z}} \\
\r_{A_2A_1}=&\,\frac{a}{3}\hat{\textbf{x}}+\frac{a}{\sqrt{3}}\hat{\textbf{y}}+\xi\, \hat{\textbf{z}} .
\end{align*}
The spin Hamiltonian for the AB$'$-stacked bilayer CrI$_3$ is reported in \eq{eq:bilayer_CrI3_Ham}, \eq{eq:CrI3_monlayer_Ham}, and \eq{eq:H_int_CrI3_AF} of the main text. While the easy axis anisotropy terms in  \eq{eq:CrI3_monlayer_Ham} favor all the spins in each layer to align along the $z$-axis, the inter-layer Hamiltonian in \eq{eq:H_int_CrI3_AF} stabilizes a anti-parallel alignment between the two layers, resulting in an anti-ferromagnetic ordering shown in \subfig{Fig:CrI3}{i}. The HP-transformation due to this AFM order, for the spins of bottom layer are
\begin{align*}
\begin{split}
  \S^{\scriptscriptstyle (A_1)-}_i=& \;(2S-a^\dagger_{1i} a_{1i})^{\frac{1}{2}}a^\dagger_{1i},\\
   \S^{\scriptscriptstyle (A_1)+}_i=&\;a_{1i}(2S-a^\dagger_{1i} a_{1i})^{\frac{1}{2}},\\
   \S^{\scriptscriptstyle (A_1)z}_i=&\;S-a^\dagger_{1i} a_{1i},
\end{split}
\Bigg|
\begin{split}   
   \S^{\scriptscriptstyle (B_1)-}_i=& \;(2S-b^\dagger_{1i} b_{1i})^{\frac{1}{2}}b^\dagger_{1i},\\
   \S^{\scriptscriptstyle (B_1)+}_i=&\;b_{1i}(2S-b^\dagger_{1i} b_{1i})^{\frac{1}{2}},\\
   \S^{\scriptscriptstyle (B_1)z}_i=&\;S-b^\dagger_{1i} b_{1i},
\end{split}
\end{align*}
and for the spins of the top layer  are 
\begin{align*}
\begin{split}
  \S^{\scriptscriptstyle (A_2)-}_i=& \;a_{2i}(2S-a^\dagger_i a_{2i})^{\frac{1}{2}},\\
   \S^{\scriptscriptstyle (A_2)+}_i=&\;(2S-a^\dagger_{2i} a_{2i})^{\frac{1}{2}}a^\dagger_{2i},\\
   \S^{\scriptscriptstyle (A_2)z}_i=&\;-S+a^\dagger_{2i} a_{2i},
\end{split}
\Bigg|
\begin{split}   
   \S^{\scriptscriptstyle (B_2)-}_i=& \;b_{2i}(2S-b^\dagger_{2i} b_{2i})^{\frac{1}{2}},\\
   \S^{\scriptscriptstyle (B_2)+}_i=&\;(2S-b^\dagger_{2i} b_{2i})^{\frac{1}{2}}b^\dagger_{2i},\\
   \S^{\scriptscriptstyle (B_2)z}_i=&\;-S+b^\dagger_{2i} b_{2i}.
\end{split}    
\end{align*}
We substitute the above into the spin Hamiltonian \eq{eq:bilayer_CrI3_Ham}, \eq{eq:CrI3_monlayer_Ham}, and \eq{eq:H_int_CrI3_AF} and use the Fourier-transformed bosonic operators (see \eq{eq:Fourier_transform_app}) to obtain the magnon Hamiltonian for the AB$'$ stacked phase  
\begin{equation}
H_\text{magnon}=\frac{1}{2}\sum\limits_{\k}\, \bm{\psi}^\dagger_\k\, \H_\k\bm{\psi}_{k}~~~\text{with}~~~\bm{\psi}_\k\equiv [\bm{\phi}_{\k}, \bm{\phi}^\dagger_{\tminus\k}]^T, \label{eq:H_mag_AFM}    
\end{equation}
where the vector $\bm{\phi}_\k$ is the same as that defined for the AB-stacked FM phase (see \eq{eq:app_BCS_magH}). The Bloch-Hamiltonian $\H_\k$ in \eq{eq:H_mag_AFM} has the BdG form \cite{Kondo2020Non};
\begin{equation}
\H_\k=\bigg[\begin{array}{cc}
        h_\k &\Delta_\k\\
        \Delta^*_{\tminus\k}&h^*_{\tminus \k}
    \end{array}\bigg]~~~~\text{with}\label{eq:app_ABpS_mag_H}
\end{equation}
\begin{align*}
h_\k=&\begin{bmatrix}
d_{1}&-\frac{\J_1}{2}\gamma^{}_{\k}&0&0\\
-\frac{\J_1}{2}\gamma^{*}_{\k}&d_{1}&0&0\\
0&0&d_{2}&-\frac{\J_2}{2} \gamma_{\k}\\
0&0&-\frac{\J_2}{2} \gamma^{*}_{\k}&d_{2}
\end{bmatrix},~~~\text{and}\\
\Delta_\k=&\;\J_\perp\left[\!\!\begin{array}{cccc}
0&0& e^{i(\frac{k_x}{3}\textbf{+}\frac{k_y}{\sqrt{3}})}&e^{i(\textbf{-}\frac{k_x}{6}\text{+}\frac{k_y}{2\sqrt{3}})}\\
0&0&e^{i\frac{k_x}{3}}&e^{\textbf{-}i(\frac{k_x}{6}\text{+}\frac{k_y}{2\sqrt{3}})} \\
e^{\textbf{-}i(\frac{k_x}{3}\textbf{+}\frac{k_y}{\sqrt{3}})}&e^{\textbf{-}i\frac{k_x}{3}}&0&0\\
 e^{i(\frac{k_x}{6}\tminus\frac{k_y}{2\sqrt{3}})}& e^{i(\frac{k_x}{6}\text{+}\frac{k_y}{2\sqrt{3}})}&0&0
\end{array}\!\!\right]
\end{align*}
where $d_{l}=\frac{3}{2}\J_l+\frac{3}{2}\lambda_l+2\kappa_l+2\J_\perp$, for $l=1,2$ and $\gamma_\k$ is defined below \eq{eq:app_ABS_coe_mag_H}. The time-reversal symmetry present in the system implies the condition ${\cal T} \H_\k{\cal T}^{-1}= \H_{-\k}$ under the time-reversal operator ${\cal T}=\sigma_x\otimes \,\sigma_0\otimes \sigma_0 K$.

We present the bilinear VMT responses generated when both $\nabla T$ and $\nabla \B$ are applied simultaneously, as functions of temperature and $\J_\perp$ in \subfigs{Fig:CrI3_bilinear_VMT}{c}{d}, respectively.

\subsection{AB-stacked bilayer anti-ferromagnet}\label{app:full_AFM}
\begin{figure}[t]
\centering
\includegraphics[width=0.985\linewidth]{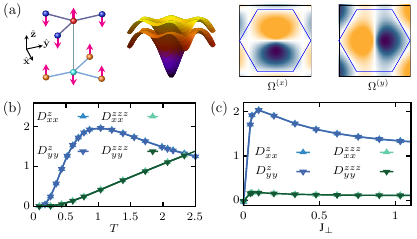}
\caption{\textbf{AB-stacked bilayer Anti-ferromagnet}. Panel (a) shows spin ordering, typical magnon-bands and the two hidden (in-plane) components of magnon Berry-curvatures, $\Omega^{(x)}$ and $\Omega^{(y)}$ for some representative parameters. The two lowest bands and the upper two bands are degenerate. The magnon-BCs are plotted in the units of $\xi$ and the yellow(blue) region represent positive (negative) values.  First Brillouin zone is
marked by a regular hexagon. 
The linear VMT responses $\sigma^{z}_{x,y}$ and $\sigma^{zz}_{x,y}$ vanish due to time-reversal symmetry. The three-fold rotation symmetry C$_3$ further forces $D^{z}_{xy}$ and $D^{zzz}_{xy}$ to vanish and constrains the remaining components to satisfy $D^{z}_{xx} = D^{z}_{yy}$ and $D^{zzz}_{xx} = D^{zzz}_{yy}$. Panel (b) shows the non-linear VMT responses $D^{z}_{xx}$, $D^{z}_{yy}$, $D^{zzz}_{xx}$, and $D^{zzz}_{yy}$ as functions of temperature $T$ for $\J_{\perp}=0.6$. Panel (c) displays the same responses as functions of the inter-layer exchange coupling $\J_{\perp}$ at fixed temperature $T=0.6$. The remaining parameters are held fixed at $\J_1=\J_2=2.2$, $\lambda_1=\lambda_2=0.09$, and $\kappa_1=\kappa_2=0.01$.
}
    \label{Fig:AB_stakced_AFM}
\end{figure}

We have seen that the vertical responses in the AB$'$-stacked AFM phase for CrI$_3$ are an order of magnitude smaller to those in the AB-stacked FM phase (compare \subfigs{Fig:CrI3}{f}{g} with \subfigs{Fig:CrI3}{l}{m}, respectively). The reason for these reduced VMT magnitudes have already been discussed in the main text. 
Now in this appendix,  we develop an alternative AB-stacked model on the bilayer honeycomb lattice that hosts anti-ferromagnetic order both within and between the layers (see \fig{Fig:AB_stakced_AFM}). We analyze the vertical transport responses for this new model (AB-stacked bilayer AFM) and compare them to the FM phase of CrI$_3$. The spin Hamiltonian for this model is given by \eq{eq:bilayer_CrI3_Ham}, \eq{eq:CrI3_monlayer_Ham}, and \eq{eq:H_int_CrI3_F}  with $\J_{1,2},\J_\perp\leq 0$.
The HP-transformations for spins at sites $A_{1,2}$ and $B_{1,2}$ for the said AFM order are given by
\begin{align*}
\begin{split}
  \S^{\scriptscriptstyle (A_l)-}_i=& \;(2S-a^\dagger_{li} a_{li})^{\frac{1}{2}}a^\dagger_{li},\\
   \S^{\scriptscriptstyle (A_l)+}_i=&\;a_{li}(2S-a^\dagger_{li} a_{li})^{\frac{1}{2}},\\
   \S^{\scriptscriptstyle (A_l)z}_i=&\;S-a^\dagger_{li} a_{li},
\end{split}
\Bigg|
\begin{split}   
   \S^{\scriptscriptstyle (B_l)-}_i=& \;b_{li}(2S-b^\dagger_{li} b_{li})^{\frac{1}{2}},\\
   \S^{\scriptscriptstyle (B_l)+}_i=&\;(2S-b^\dagger_{li} b_{li})^{\frac{1}{2}}b^\dagger_{li},\\
   \S^{\scriptscriptstyle (B_l)z}_i=&\;-S+b^\dagger_{li} b_{li},
\end{split}
\end{align*}
with $l=1,2$. The Fourier transformed boson operators for this model are also defined by \eq{eq:Fourier_transform_app}. The resulting magnon Hamiltonian is similar to those of AB$'$-stacked AFM phase given in \eq{eq:H_mag_AFM} 
with 
\begin{align}
h_\k=&\begin{bmatrix}
d_{1}&0&0&0\\
0&d_{1}+\J_\perp&0&0\\
0&0&d_{2}+\J_\perp&0\\
0&0&0&d_{2}
\end{bmatrix},~~~\text{and}\notag\\
\Delta_\k=&\;\J_\perp\left[\!\!\begin{array}{cccc}
0&\frac{\J_1}{2}\gamma^{}_{\k}& 0&0\\[-2pt]
\frac{\J_1}{2}\gamma^{*}_{\k}&0&\J_\perp&0 \\[-2pt]
0&\J_\perp&0&\frac{\J_2}{2} \gamma_{\k}\\[-2pt]
0&0&\frac{\J_2}{2} \gamma^{*}_{\k}&0
\end{array}\!\!\right]\notag
\end{align}
where $d_{l}=\frac{3}{2}\J_l+\frac{3}{2}\lambda_l+2\kappa_l$, for $l=1,2$ and $\gamma_\k$ is the same as defined for the ferromagnetic case (see below \eq{eq:app_ABS_coe_mag_H}).
The system shows time-reversal symmetry in the single magnon sector described by the same time-reversal operator as the AB$'$-stacked AFM phase (see below \eq{eq:app_ABpS_mag_H}).
Similar to the AB-stacked FM phase of CrI$_3$, this model possesses reflection symmetry $k_x\rightarrow -k_x$, as well as the three fold rotation symmetry C$_3$ about the $z$-axis. We numerically diagonalize the magnon Bloch Hamiltonian of this model using paraunitary matrices to obtain the four independent magnon modes. We plot magnon band dispersions in \subfig{Fig:AB_stakced_AFM}{a}(inset) for some typical parameter values. The two lowest bands and the upper two bands are
degenerate as we have set $\delta=\J_2/\J_1=\lambda_2/\lambda_1=\kappa_2/\kappa_1=1$.
The pseudo-$\Z$ operator is same as AB-stacked FM phase of CrI$_3$. We plot the two components of the hidden Berry curvature, $\Omega^{(x)}$ and $\Omega^{(y)}$ for the lowest energy magnon modes in \subfig{Fig:AB_stakced_AFM}{a}(inset).

The linear  VMT responses, $\sigma^z_{x,y}$ and $\sigma^{zz}_{x,y}$ vanish due to presence of time-reversal symmetry.  The three fold rotation symmetry  C$_3$ makes the responses $D^{z}_{xy}, D^{zzz}_{xy}$ to vanish and constrains remaining responses to follow  $D^{z}_{xx}=D^{z}_{yy}$ and $D^{zzz}_{xx}=D^{zzz}_{yy}$. We plot the non-linear VMT responses $D^{z}_{xx},D^{z}_{yy},D^{zzz}_{xx}$ and $D^z_{yy}$ as a function of temperature $T$ for $\J_{\perp}=0.6$  in \subfig{Fig:AB_stakced_AFM}{b} and as a function of inter-layer coupling $\J_\perp$ in \subfig{Fig:AB_stakced_AFM}{c} for $T=0.6$. The other parameters are fixed at $\J_1=\J_2=2.2,\lambda_1=\lambda_2=0.09 $ and $\kappa_1=\kappa_2=0.01$.  
\bibliographystyle{unsrt}
\bibliography{ref.bib}
\end{document}